\documentclass[10pt,journal,compsoc]{IEEEtran}
\usepackage{subfigure}
\usepackage{graphicx}
\usepackage{array}
\usepackage{graphicx}
\usepackage{xcolor}
\usepackage{amsmath}
\usepackage{multirow}
\usepackage{ragged2e}
\usepackage{subfig}
\usepackage{ulem}
\usepackage{rotating}
\usepackage[ruled,linesnumbered]{algorithm2e}
\ifCLASSOPTIONcompsoc
  \usepackage[nocompress]{cite}
\else
  \usepackage{cite}
\fi

\begin{document}

\title{On the Merge of \textit{k}-NN Graph}

\author{Wan-Lei Zhao*, Hui Wang, Peng-Cheng Lin
                and Chong-Wah Ngo,~\IEEEmembership{Senior Member,~IEEE}
\IEEEcompsocitemizethanks{\IEEEcompsocthanksitem Wan-Lei Zhao, Hui Wang and Peng-Cheng Lin are with the Department of Computer Science, Xiamen University, Xiamen, China. Wan-Lei Zhao is the corresponding author.
E-mail: wlzhao@xmu.edu.cn}
\IEEEcompsocitemizethanks{\IEEEcompsocthanksitem Chong-Wah Ngo is with School of Computing and Information Systems,
Singapore Management University. }
\thanks{Manuscript received April 8, 2020; revised xxx, 2020.}}

\IEEEtitleabstractindextext{
\begin{justify}
\begin{abstract}
\textit{k}-nearest neighbor graph is a fundamental data structure in many disciplines such as information retrieval, data-mining, pattern recognition, and machine learning, etc. In the literature, considerable research has been focusing on how to efficiently build an approximate \textit{k}-nearest neighbor graph (\textit{k}-NN graph) for a fixed dataset. Unfortunately, a closely related issue of how to merge two existing \textit{k}-NN graphs has been overlooked. In this paper, we address the issue of \textit{k}-NN graph merging in two different  scenarios. In the first scenario, a symmetric merge algorithm is proposed to combine two approximate \textit{k}-NN graphs. The algorithm facilitates large-scale processing by the efficient merging of \textit{k}-NN graphs that are produced in parallel. In the second scenario, a joint merge algorithm is proposed to expand an existing \textit{k}-NN graph with a raw dataset. The algorithm enables the incremental construction of a hierarchical approximate \textit{k}-NN graph. Superior performance is attained when leveraging the hierarchy for NN search of various data types, dimensionality, and distance measures.

\end{abstract}
\end{justify}

}

\maketitle

\IEEEdisplaynontitleabstractindextext

\IEEEpeerreviewmaketitle

\section{Introduction}
\label{sec:introduction}
Given a dataset $S=\{ x{\mid}x \in R^d\}$ with \textit{n} samples, \textit{k}-NN graph refers to the data structure, $G[i]$, that keeps the top-\textit{k} nearest neighbors of every sample $x_i$ in the dataset. It is a key data structure in the manifold learning~\cite{isomap, lle}, data mining, machine learning~\cite{mlsurvey20}, and information retrieval. It is also supportive for balanced graph construction~\cite{uai12:jwang}. Basically, given a metric $m(\cdot,\cdot)$, the construction of \textit{k}-NN graph is to find the top-\textit{k} nearest neighbors for each data sample. When being built in a brute-force way, the time complexity is $O(d{\cdot}n^2)$, where \textit{d} is the dimension and \textit{n} is the size of a dataset. As the value of \textit{n} can be very large in big data applications, the construction of an exact \textit{k}-NN graph is computationally expensive. The problem is even challenging if the value of \textit{d} is also large, \textit{i.e.}, the samples are in high dimensional space. Due to its high time complexity, the existing works~\cite{weidong, efanna16, jmlr09, msraKNN} instead aim for an efficient but approximate solution. 

Despite constant progress has been made in the last few decades, to the best of our knowledge, there is no research dedicated for merging \textit{k}-NN graphs. Given two \textit{k}-NN graphs $G_1$ and $G_2$ from the datasets $S_1=\{ x{\mid}x \in R^d\}$ and $S_2 = \{ y{\mid}y \in R^d\}$ respectively, the task of merging is to build $G$ for $S=S_1 \cup S_2$ based on $G_1$ and $G_2$ instead of reconstructing $G$ from scratch\footnote{Without the loss of generality, we assume $S_1 \cap S_2 = {\emptyset}$.}. During merging, $G_1$ is updated with the nearest neighbor samples discovered on $S_2$. A similar process is carried out to update $G_2$ with $S_1$.

The problem is important for enlightening large-scale construction of approximate \textit{k}-NN graph in an incremental manner. For a fixed dataset, merging ensures an efficient divide-and-conquer way of graph construction. In some applications, \textit{k}-NN graphs can only be built for individual subsets on the initial stages, given the data are distributed on different nodes or not all of them are ready at that moment. In the case of parallel computing, one would prefer slicing the data into blocks and computing the approximate \textit{k}-NN graph for each block on different machines. The subgraphs are later reduced into one by repeatedly merging two subgraphs each time.
For the online data streamed in batch, merging is a fundamental step to expand an already-built \textit{k}-NN graph with new raw samples. The step naturally extends an approximate \textit{k}-NN graph construction algorithm to online version. The typical scenario takes place in a multimedia website such as Flickr, where the \textit{k}-NN graph is built to support content-based photo search. In this scenario, the constructed \textit{k}-NN graph should be merged with new nodes when new photos are uploaded from time to time. Considering the big number of photos and the high upload frequency in the website, the re-construction of the graph from scratch from time to time is nearly impossible. This is where the \textit{k}-NN graph merge algorithm could fit in.

A brute force way of merging graph $G_1$ with graph $G_2$ can be undertaken by a thorough cross-comparison between samples in $S_1$ and samples in $S_2$. The produced edge, say $<x_i, y_j, m(x_i, y_j)>$, is inserted into $G_1[i]$ and $G_2[j]$ if it is ranked at top-\textit{k} of the corresponding NN list. Given the cardinalities of $S_1$ and $S_2$ are $n_1$ and $n_2$ respectively, the time complexity of such cross-comparison is $O({n_1}{\cdot}{n_2}{\cdot}d)$,  which is even higher than constructing the graph from scratch by algorithms such as~\cite{weidong, efanna16}. Hence, the problem of merging is not trivial.

In this paper, two efficient and generic algorithms for \textit{k}-NN graph merge are proposed. They are designed to address the merging issues under two different scenarios. 
In the first scenario, we address the issue of merging two already-built \textit{k}-NN graphs. Merging more than two graphs can be trivially implemented with a recursive call of two-graph merging. In the second scenario, we address the issue of merging a raw sample set into an already built \textit{k}-NN graph. The merging algorithm novelly leads to a hierarchical approximate \textit{k}-NN graph construction algorithm. Capitalizing on the hierarchical structure, an efficient nearest neighbor (NN) search algorithm is presented.  

The remainder of this paper is organized as follows. In Section~\ref{sec:rela}, a brief review of the research works on approximate \textit{k}-NN graph construction and NN search is presented. Section~\ref{sec:merge} presents two algorithms for \textit{k}-NN graph merge under two different scenarios. They are called symmetric merge (S-Merge) and joint merge (J-Merge). Deriving from J-Merge, a hierarchical approximate \textit{k}-NN graph construction algorithm is presented at the end of this section. In Section~\ref{sec:srch}, based on the constructed hierarchical approximate \textit{k}-NN graphs, an efficient NN search approach is presented. Section~\ref{sec:exp} presents empirical findings to justify the effectiveness of the proposed algorithms for \textit{k}-NN graph merging and NN search. Section~\ref{sec:conc} concludes the paper.

\section{Related Works}
\label{sec:rela}
\subsection{Approximate \textit{k}-NN Graph Construction}
In the approximate \textit{k}-NN graph construction, there are in general two types of approaches. The space partitioning approaches such as~\cite{jmlr09,msraKNN,ecml13:zhang} follow a two-stage pipeline. At the first stage, samples are divided into a number of small subsets by random partitions. The random partitions over the space can be generated using techniques such as recursive Lanczos bisection~\cite{jmlr09}, hierarchical random projections~\cite{msraKNN}, or a series of locality-sensitive hash functions~\cite{ecml13:zhang}. The samples in each subset are expected to be close to each other. Exhaustive pairwise comparisons are carried out within each subset. The closeness relations (\textit{viz.}, edges in the \textit{k}-NN graph) between any two samples in one subset are established. Based on the closeness relations, the second stage selects samples to update the \textit{k}-NN graph. Specifically, these samples substitute the existing samples with longer edges in the \textit{k}-NN graph. The two stages iterate and perform updates incrementally. Due to the difficulty in designing partitioning schemes generic for various metric spaces, this type of approach is generally effective in $\textit{l}_p$-space only.

NN-Descent~\cite{weidong}, the second type of \textit{k}-NN graph construction, differs by its ability to operate under various distance measures. The algorithm starts the construction from a random \textit{k}-NN graph. Based on the principle of ``neighbor's neighbor is likely to be the neighbor'', the cross-comparison is invoked for the samples staying in a neighborhood region. The closer pairs of samples are then selected to update the neighborhood of a target sample. The iteration continues until no change of neighborhood configuration. As the algorithm essentially performs hill-climbing optimization batchfully~\cite{icai11:kiana,weidong} independent of metric space, it is both generic and efficient.
Recently, the hybrid scheme based on both the space partitioning approach and NN-Descent has been proposed in the literature~\cite{efanna16}. Although efficient, this hybrid scheme is infeasible for the metrics beyond $l_p$-norms.

Other efficient approaches include KIFF~\cite{icde16:boutet} which is specifically designed to build \textit{k}-NN graph for the high-dimensional sparse dataset. The \textit{k}-NN graph is built by sequentially considering the sample pairs with high co-occurrence frequency. This approach turns out to be much more efficient than NN-Descent~\cite{weidong} on sparse datasets.

All the aforementioned approaches assume \textit{k}-NN graph construction under static dataset. To cope with streaming data, such as photos and videos uploaded on a daily basis to the social media platform, an incremental update is demanded. Nevertheless, there is no such incremental updating strategy being considered in these algorithms. Although NN-Descent can be applied to augment an already-built \textit{k}-NN graph with new samples, the data aggregation cost is high if updates are performed frequently.

Online graph construction is proposed in~\cite{infosys13:yury,pami18:yury} primarily for the nearest neighbor search. These approaches, nevertheless, are not designed to build a \textit{k}-NN graph. In order to increase search efficiency, the \textit{k}-NN graph is sparsified during the construction. Consequently, the right samples supposedly to be included in the \textit{k}-NN list are deliberately omitted when being occluded by other neighboring samples. Instead, the links to remote neighbors are maintained. During the search, these links serve as short cuts to allow quick access to the close neighborhood of a query. However, due to the amendment, the graph is no longer a \textit{k}-NN graph by definition and does not support tasks beyond NN search.

This paper addresses the problem of merging \textit{k}-NN graphs in two typical scenarios. In our solution, the merge operation enables the dynamic construction of \textit{k}-NN graph, through the incremental update of the graph with the raw dataset. Moreover, like NN-Descent~\cite{weidong}, our algorithm is generic to various distance measures.

\subsection{NN Search}
An issue that is closely related to approximate \textit{k}-NN graph construction is the nearest neighbor search. The primary goal of NN search is to find out the nearest neighbors from a given dataset for a query sample. In the problem, both the query and the candidate samples are assumed to be from the same space \textit{i.e.}, $R^d$.

This issue has been traditionally addressed by a variety of tree partitioning approaches, such as K-D tree~\cite{kdtree75}, R-tree~\cite{rtree84}, X-Tree~\cite{xtree96}, NV-tree~\cite{pami09:lejsek} and Trinary-Projection tree~\cite{pami14:wang}. The idea is to partition a space into hierarchical sub-spaces such that only a few branches of sub-spaces are traversed during NN search. However, unlike B-tree in 1D case, the true nearest neighbor may reside in the branches that are outside the candidate sets. Therefore, extensive probing over a large number of branches in the tree becomes inevitable. Recent indexing structures FLANN~\cite{pami14:flann} and Annoy~\cite{cvpr08:annoy} partition the space with hierarchical \textit{k}-means and multiple K-D trees respectively. Although both of them are efficient, sub-optimal results are achieved.

Apart from tree partitioning approaches, the quantization based approaches~\cite{ivfrvq10, JDS11,artem16,icml14:tzhang} and locality-sensitive hashing (LSH)~\cite{lsh04, mlsh07, cvpr12:SupHash, TSCG04:Locality, pami18:wang} have been extensively explored in the last decade. Compared to tree partitioning, these approaches are efficient in terms of both memory and speed. The speedup is significant especially when the size of encoding bits is short. Nevertheless, these advantages are traded off by relatively lower search accuracy. In addition, these approaches are mostly only suitable for $l_p$-norms. The design of generic hash functions is non-trivial.

Recently, the graph-based approaches such as hill-climbing~\cite{icai11:kiana} and nearest neighbor descent (NN-Descent)~\cite{weidong}, demonstrate superior performance over other categories of approaches in many large-scale NN search tasks~\cite{mm12:jdwang,pami18:yury, dpg:wenli,olg:wlzhao}. All the approaches in this category are built upon an approximate \textit{k}-NN graph or diversified approximate \textit{k}-NN graph. The search procedure starts from a group of random seeds and traverses iteratively over the graph by the best-first search. Guided by the neighbors of visited vertices, the search procedure descents closer to the true nearest neighbor in each round until no better candidates could be found. The approaches in~\cite{icai11:kiana,pami18:yury,infosys13:yury,efanna16,olg:wlzhao, nsg19,pr19:hcnng} in general follow a similar search procedure. The major difference between them lies in the structure of graphs upon which the NN search is undertaken. For most of the graph-based approaches, the space complexity is roughly linear to the scale of a dataset. And the extra merit is that they are suitable for various distance measures.

In this paper, a hierarchical approximate \textit{k}-NN graph construction algorithm, as a by-product of graph merging operation, is derived for NN search. Similar to HNSW~\cite{pami18:yury}, the hierarchical graph achieves speedup by skipping samples lying in the far neighborhood of a query. The speedup is particularly significant in low dimensional space. Compared to HNSW, the algorithm offers relatively better search performance due to the high quality of the approximate \textit{k}-NN graph. Furthermore, as the intermediate result of graph construction, the hierarchical structure is not required to be in memory during construction. The burden in maintaining and updating the hierarchy is therefore relieved. Moreover different from HNSW, the produced graph structure also facilitates other tasks such as browsing over close neighbors of each sample (e.g., photos) since it maintains an approximate \textit{k}-NN graph.

\section{Two \textit{k}-NN Graph Merge Strategies}
\label{sec:merge}
In this section, two schemes that are used to merge the \textit{k}-NN graphs in two different scenarios are developed. In the first scenario, it is assumed that two sub-\textit{k}-NN graphs are already built by any existing algorithms such as~\cite{weidong, efanna16, olg:wlzhao} or in a brute-force way. The algorithm merges two subgraphs into one. In the second scenario, the merge algorithm deals with the problem of joining a raw set into an already built \textit{k}-NN graph. Without loss of generality, we assume that there is no intersection between the two subsets to be merged in both scenarios. Finally, a hierarchical approximate \textit{k}-NN graph construction algorithm is proposed as a derivation from the second merge scheme.
\begin{figure*}
\begin{center}
\subfigure[two \textit{2}-NN graphs]
  {\includegraphics[width=0.16\linewidth]{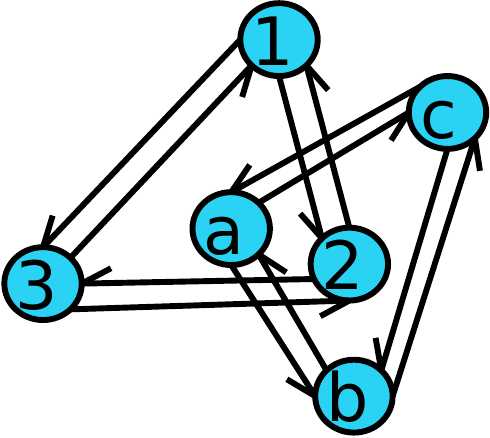}}
  \hspace{0.1in}
 \subfigure[merge operations]
  {\includegraphics[width=0.52\linewidth]{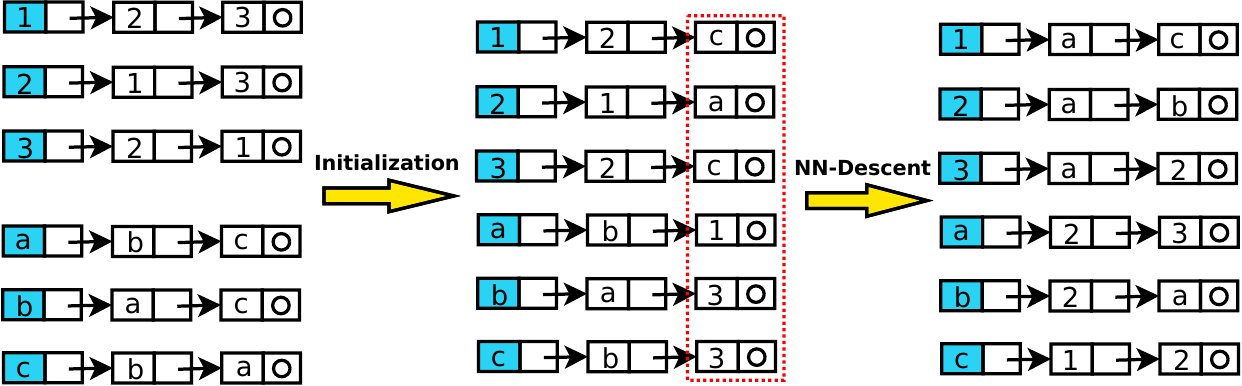}}
  \hspace{0.1in}  
  \subfigure[result]
  {\includegraphics[width=0.16\linewidth]{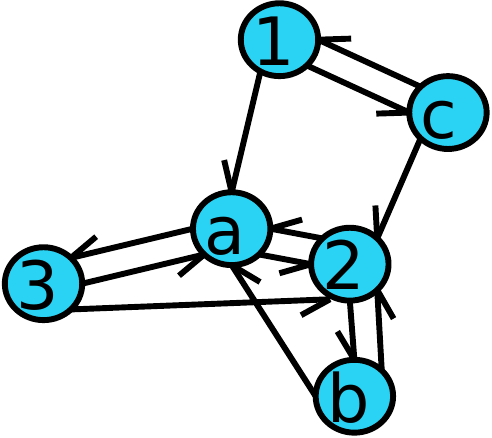}}  
  \caption{The illustration of symmetric merge (S-Merge). Figure (a) shows two disconnected subgraphs. For distinction, the vertices of two subgraphs are denoted as numbers $S_1=\{1,2,3\}$ and alphabets $S_2=\{a,b,c\}$ respectively. Each subgraph is a \textit{2}-NN graph. Figure (b) illustrates the three steps in S-Merge. Figure (c) is the resulting graph after S-Merge.}
\label{fig:merge_strategy}
\end{center}
\end{figure*}

\subsection{Symmetric Merge}
Given datasets $S_1=\{s_i|s_i \in R^d\}$ and $S_2=\{s_j|s_j \in R^d\}$ ($S_1 \cap S_2 = \emptyset$), \textit{k}-NN graphs $G_{1}$ and $G_{2}$ have been built for $S_1$ and $S_2$ respectively. Now we consider the problem of constructing \textit{k}-NN graph $G$ for dataset $S=S_1 \cup S_2$ based on $G_{1}$ and $G_{2}$. In our solution, we propose to merge the graphs in three major steps, which are detailed in Alg.~\ref{alg:pmerge}.

\begin{algorithm}
    \KwData{$S$: dataset; $G_{1}$: \textit{k}-NN Graph of $S_1$ of size \textit{m}; $G_{2}$:  \textit{k}-NN Graph of $S_2$ of size \textit{n}}
    \KwResult{$G$: \textit{k}-NN Graph $G$}
    \textbf{Divide} $G_{1}$ into $G^{u}_{1}$ and $G^{v}_{1}$\;
    \textbf{Divide} $G_{2}$ into $G^{u}_{2}$ and $G^{v}_{2}$\;
    \For{i=1; i $\leq$ m; i++}
 	{	 
	    \textbf{Append} $G^{u}_{1}[i]$ with $\frac{k}{2}$ random samples from $S_2$\;
	}
	\For{i=1; i $\leq$ n; i++}
	{
	    \textbf{Append} $G^{u}_{2}[i]$ with $\frac{k}{2}$ random samples from $S_1$\;
	}
	$G \leftarrow G^{u}_{1} \cup G^{u}_{2} $\;

	\Repeat{$c == 0$}
	{
		$\overline{G} \leftarrow Reverse(G)$\;
		$\mathcal{G}[i] \leftarrow \overline{G}[i]\cup G[i]$, $\forall i \in G$\;
		$c \leftarrow 0$\;
		\For{$u  \in \mathcal{G}$}
		{
			\For{$s_{i},s_{j} \in \mathcal{G}[u]$,$s_{i}\neq s_{j}$}
			{
				\uIf{$s_{i} \in S_1$ \& $s_{j} \in S_2$ or $s_{i} \in S_2$ \& $s_{j} \in S_1$}
				{
				$l \leftarrow m(s_{i},s_{j})$\;
				UpdateNN$(G[s_{i}],s_{j}, l)$\;
				UpdateNN$(G[s_{j}],s_{i}, l)$\;
				$c \leftarrow c + 1$\;
				}
			}
		}
	}
	$G$ $\leftarrow$ \textbf{Merge and Sort} $G$ with $G^{v}_{1}$ and $G^{v}_{2}$ \;
	\Return $\textit{k}$-NN Graph $G$
  \caption{Symmetric Merge (S-Merge)}
 \label{alg:pmerge}
\end{algorithm}

In the first step, the rear $\frac{k}{2}$ samples in each \textit{k}-NN list from $G_{1}$ and $G_{2}$ are truncated out (\textit{Lines 1-2}). These truncated lists are kept for later use. After the truncation, there are $\frac{k}{2}$ samples in each NN list of $G_{1}$ and $G_{2}$. Each NN list of the graphs is then appended with $\frac{k}{2}$ samples randomly drawn from a different dataset. For instance, $G^{u}_{1}$ is appended with the samples from $S_2$ (\textit{Lines 3-5}) and similarly for $G^{u}_{2}$ (\textit{Lines 6-8}). Combining the augmented $G^{u}_{1}$ and $G^{u}_{2}$ leads to a half-baked \textit{k}-NN graph $G$ for the set $S=S_1 \cup S_2$ (\textit{Line 9}). In the second step, the NN-Descent iteration~\cite{weidong} is performed on each NN list of graph $G$. The iteration continues until it converges (\textit{Lines 10-23}). In \textit{Line 11}, the function $\mathit{Reverse(G)}$ returns a reverse graph for the graph $G$~\cite{weidong}, which keeps the reverse NN list for each sample. In \textit{Lines 18-19}, function $\mathit{UpdateNN(G[s_i],s_j,l)}$ inserts a new edge $<s_i, s_j, l>$ into the NN list of $s_i$. Function $\mathit{UpdateNN}$ takes the \textit{k}-NN list of $s_i$, sample ID of $s_j$ and the distance \textit{l} between $s_i$ and $s_j$ as the input parameters.

Different from~\cite{weidong}, the cross-comparison on each \textit{k}-NN list only takes place between samples from two different sets, namely between $s_i \in S_1$ and $s_j \in S_2$. Finally, the truncated rear lists from $G_{1}$ and $G_{2}$ are combined with $G$ by a simple merge sort on each \textit{k}-NN list. The top-\textit{k} samples in each NN list are kept (\textit{Line 24}) after the merge sort. Then the approximate \textit{k}-NN graph for the whole set $S$ is constructed. The merge algorithm is illustrated in Fig.~\ref{fig:merge_strategy}.

Since the merge is undertaken on two constructed sub-\textit{k}-NN graphs, the merge algorithm is called symmetric merge (S-Merge) from now on. In the first step of S-Merge, \textit{50\%} of samples in each \textit{k}-NN list of graphs $G_{1}$ and $G_{2}$ are cut out for the final merge. Whether to reserve more or fewer than \textit{50\% }of neighbors in a \textit{k}-NN list is an empirical setting. As will be shown in the ablation analysis (Section~\ref{sec:expaba}), a merged graph is maintained in high quality when the proportion of samples being kept is in the range of $[\frac{k}{5},~\frac{k}{2}]$.

S-Merge is particularly helpful for building an approximate \textit{k}-NN graph in parallel. Imagine that multiple subgraphs are built with different threads or on different nodes by any existing algorithms~\cite{weidong, olg:wlzhao}. S-Merge is repeatedly called whenever any two subgraphs are ready to be merged. The process continues until a complete \textit{k}-NN graph for the entire dataset is constructed.

\subsection{Joint Merge}
Based on a similar idea as S-Merge, the joint merge algorithm deals with the problem of merging a raw set into a built \textit{k}-NN graph. Given two datasets $S_1=\{s_i|s_i \in R^d\}$ and $S_2=\{s_j|s_j \in R^d\}$, only the \textit{k}-NN graph $G_{1}$ for $S_1$ is available. The problem is to build a graph $G$ for dataset $S=S_1 \cup S_2$. Intuitively, exhaustive comparisons could be performed between the samples from $S_1$ and $S_2$ and within $S_2$. In this way, the \textit{k}-NN lists in graph $G_{1}$ are updated while the \textit{k}-NN lists for $S_2$ are established. Unfortunately, the time complexity of this way is high. Another possible solution is to construct a \textit{k}-NN graph for $S_2$ by NN-Descent and perform S-Merge between the two subgraphs. It is feasible, whereas, not the most cost-effective solution. Following a similar idea as S-Merge, the joint merge (J-Merge) is proposed to fulfill the merge. J-Merge algorithm is summarized in Alg.~\ref{alg:jmerge}.

\begin{algorithm}
    \KwData{$S$: dataset; $G_{1}$: \textit{k}-NN Graph of $S_1$ of size \textit{m}; $S_2$: raw dataset of size \textit{n}}
    \KwResult{$G$: \textit{k}-NN Graph $G$}
    
    \textbf{Divide} $G_{1}$ into $G^{u}_{1}$ and $G^{v}_{1}$\;
    \For{i=1; i $\leq$ m; i++}
 	{	 
	    \textbf{Append} $G^{u}_{1}[i]$ with $\frac{k}{2}$ random samples from $S_2$\;
	}
	\For{i=1; i $\leq$ n; i++}
	{
	    \textbf{Initialize} $G_{2}[i]$ with $k$ random samples from $S_1 \cup S_2$\;
	}
	$G \leftarrow G^{u}_{1} \cup G_{2} $\;

	\Repeat{$c == 0$}
	{
		$\overline{G} \leftarrow Reverse(G)$\;
		$\mathcal{G}[i] \leftarrow \overline{G}[i]\cup G[i]$, $\forall i \in G$\;
		$c \leftarrow 0$\;
		\For{$u  \in \mathcal{G}$}
		{
			\For{$s_{i},s_{j} \in \mathcal{G}[u]$,$s_{i}\neq s_{j}$}
			{
				\uIf{$s_{i} \in S_1$ \& $s_{j} \in S_2$ or $s_{i} \in S_2$ \& $s_{j} \in S_1$ or $s_{i}, s_{j} \in S_2$  }
				{
				$l \leftarrow m(s_{i},s_{j})$\;
				UpdateNN$(G[s_{i}],s_{j}, l)$\;
				UpdateNN$(G[s_{j}],s_{i}, l)$\;
				$c \leftarrow c + 1$\;
				}
			}
		}
	}
	$U$ $\leftarrow$ \textbf{Merge and Sort} $G$ with $G^{v}_{1}$\;
	\Return $\textit{k}$-NN Graph $G$
  \caption{Joint Merge (J-Merge)}
 \label{alg:jmerge}
\end{algorithm}

The \textit{k}-NN graph $G_{1}$ is initially cut into $G^{u}_{1}$ and $G^{v}_{1}$. $G^{u}_{1}$ keeps the top $\frac{k}{2}$-NN lists of $G_{1}$, while $G^{v}_{1}$ maintains the remaining lists in $G_{1}$ for merge sort in the final stage. The NN lists of $G^{u}_{1}$ are appended with $\frac{k}{2}$ samples randomly drawn from $S_2$. For dataset $S_2$, a random \textit{k}-NN list is initialized for each sample.  The samples are selected from $S$ (\textit{i.e.}, the union of $S_1$ and $S_2$). This raw graph is denoted as $G_{2}$. Combining with $G^{u}_{1}$, we have a half-baked \textit{k}-NN graph $G$ for the set $S$. The iteration continues until no new update happens. Finally, the truncated $G^{v}_{1}$ is combined with $G$ by merge sort on each \textit{k}-NN list. The approximate \textit{k}-NN graph for the whole set $G$ is produced by keeping the top-\textit{k} samples in each NN list. Similar to S-Merge, the proportion of samples to be chopped from a \textit{k}-NN list of $G_{1}$ is an empirical setting. As will be verified in the experiment, the proportion in the range of $[\frac{k}{3}, \frac{k}{2}]$ suffices for a graph with consistent performance. Without loss of generality, each \textit{k}-NN list in $G_{1}$ is cut into equal halves by default. This setting remains valid as the scales of the two subsets are unbalanced.

Both S-Merge and J-Merge algorithms spend most of the costs on the distance computation (\textit{Line 17} in Alg.~1 and \textit{Line 16} in Alg.~2). The time costs that are spent on merge operation with the truncated graph (\textit{Line 24} in Alg.~1 and \textit{Line 23} in Alg.~2) are negligible compared to the cost on the 3rd step when the data dimensionality is high. Compared to S-Merge, J-Merge needs to perform cross-comparison between samples within the raw set. As a result, for a dataset of the same size, J-Merge requires more comparisons than that of S-Merge.

\subsection{Hierarchical \textit{k}-NN graph Construction via J-Merge}
Different from S-Merge and NN-Descent, J-Merge is capable of building an approximate \textit{k}-NN graph for a dataset where its size grows incrementally. Specifically, the construction of an approximate \textit{k}-NN graph starts when only a small subset of a dataset is available. By using J-Merge, the remaining subsets are incrementally combined with the increasingly growing \textit{k}-NN graph. Therefore, J-Merge indeed can be applied as an incremental version of \textit{k}-NN graph construction. The size of a subset to be joined in can be specified. Typically, when the size is equal to the current graph, a newly updated \textit{k}-NN graph will double its size after merging.

Namely, the construction starts by building an approximate \textit{k}-NN graph on a randomly sampled small subset from dataset $S$ with NN-Descent. Thereafter, a raw subset of the same size as the baked one is sampled from the remaining dataset of $S$. J-Merge is called to join this subset into the approximate \textit{k}-NN graph. After J-Merge, the \textit{k}-NN graph grows two times bigger than before. In the next round of merge, the size of the block to be joined in is two times bigger than the previous round. This sampling and merging process is repeated for several rounds until all the samples in $S$ are joined into the graph. The intermediate \textit{k}-NN graphs produced during the whole iterations form a hierarchical (or pyramid) structure. The lower the layer is, the more samples are kept in the graph. The bottom layer is the approximate \textit{k}-NN graph built for the whole set $S$. This approximate \textit{k}-NN construction algorithm is called hierarchical merge (H-Merge). Essentially, H-Merge is a repetitive calling of J-Merge. It is important that the subset joined in the hierarchy each time should be randomly drawn from the whole set. This is to guarantee that different levels in the hierarchy reflect well the data distribution of different granularity.

Similar to the online \textit{k}-NN graph construction algorithm~\cite{olg:wlzhao}, H-Merge deals with the \textit{k}-NN graph construction problem for a close set as well as an open set. However, different from~\cite{olg:wlzhao}, the \textit{k}-NN graph is constructed in a hierarchical manner. Compared to the hierarchical navigable small world (HNSW) graphs, H-Merge produces an approximate \textit{k}-NN graph instead of a sparsified approximate \textit{k}-NN graph in each layer. Furthermore, the hierarchy in H-Merge is formed by intermediate \textit{k}-NN graphs during the merge. No real hierarchical structure is maintained during the construction. The approximate \textit{k}-NN graph of one layer is derived from the approximate \textit{k}-NN graph of the upper layer. The upper layer graph is merged into the approximate \textit{k}-NN graph of the next layer after each round of J-Merge. During the construction process, one is allowed to save an arbitrary number of layers to form the hierarchy.

Additionally, similar to HNSW graphs, the hierarchical structure formed by H-Merge could be adopted for the nearest neighbor search. Like HNSW graphs, the links to the remote neighbors are kept in the upper layers of the hierarchy. These remote links help to skip a large number of far neighbors during the NN search. As a result, a higher speedup over NN search on a single flat graph is expected, particularly on low dimensional data.  These remote links also act as the bridges between clustered data such that the NN search escapes from being trapped in the clustered local.  Furthermore, due to the high quality of the approximate \textit{k}-NN graph, the NN search performance based on such hierarchy turns out to be even better than HNSW graphs in some cases, which will be revealed in the experiment section. In order to make better use of the hierarchical graphs produced by H-Merge, the graph in each hierarchy is undergone graph diversification before the hierarchical graphs are supplied to support NN search. This will be detailed in Section~\ref{sec:srch}.

There is one parameter in the above merge algorithms, namely parameter \textit{k}, which is also the size of the resulting \textit{k}-NN graph. For S-Merge and J-Merge, \textit{k} is specified as the user's wishes. In general, larger \textit{k} leads to better \textit{k}-NN graph quality while higher computation cost. Besides \textit{k}, the \textit{k}-NN graph quality is also impacted by the quality of the initial sub-\textit{k}-NN graphs in both scenarios.
In H-Merge, parameter \textit{k} has a similar impact on the performance. Since the scale of approximate \textit{k}-NN graphs of the non-bottom layer is small. It is no need to keep a full \textit{k}-NN list for these graphs. As a result, on these layers, the size of the \textit{k}-NN list is set to $\frac{k}{2}$.

Essentially, S-Merge, J-Merge, and H-Merge are the extensions of NN-Descent algorithm. S-Merge is designed for parallel approximate \textit{k}-NN graph construction. J-Merge is designed to build an approximate \textit{k}-NN graph dynamically. While H-Merge is proposed to support fast NN search for an open set. All the above three graph merge algorithms are generic to various distance metrics. The codes of all these algorithms have been open-sourced on GitHub\footnote{https://github.com/wlzhao/nn-merge}.

\subsection{Convergence, Optimality and Complexity Analysis}
\label{sec:conv}
In this section, the convergence analysis is made for S-Merge. Since the optimization strategies used in S-Merge, J-Merge, and H-Merge are similar. The convergence analysis for S-Merge is also feasible for J-Merge, H-Merge as well as NN-Descent. 

Given datasets $S_1=\{s_i|s_i \in R^d\}$, $S_2=\{s_j|s_j \in R^d\}$ and $S=S_1 \cup S_2$, $U^0$ is the union of augmented graphs $G^u_{1}$ and $G^u_{2}$, which is the initial graph prepared for NN-Descent iteration. Given a \textit{k}-NN graph $G$, we define function $\mathit{\phi(G)}$ that returns the sum of distances from all \textit{k} neighbors of all samples kept in $G$.
\begin{equation}
\phi(G)=\sum_{i=1}^{n}\sum_{j=1}^{k}G_{ij},
\label{eqn:sum}
\end{equation}
where $G_{ij}$ keeps the distance from the $j$-th neighbor to sample $i$. The set of the possible states for the approximate \textit{k}-NN graph is finite and countable. Its size is $n{\times}C_n^k$, where \textit{n} is the input dataset size and \textit{k} is the size of each NN list. Essentially, NN-Descent simply jumps from one state (solution) to another more optimal one.
Given the true \textit{k}-NN graph for dataset $S$ is $\mathcal{G}$, we have $\phi(U^0) > \phi(\mathcal{G})$ holds\footnote{Without the loss of generality, the shorter the distance is, the closer is the neighbor.}. After one round iteration in Alg.~\ref{alg:pmerge}, closer neighbors are joined into NN lists and far neighbors are swapped out. To make the update operation happens, there must exist a $U^{0}_{ij}$ being replaced by $U^{1}_{ij}$, where $U^{1}_{ij} < U^{0}_{ij}$. So we have $\phi(U^0) > \phi(U^1)$. Given the series of intermediate \textit{k}-NN graphs produced by the procedure are $U^1, U^2, \cdots, U^t,\cdots$, following inequation holds
\begin{equation}
\phi(U^0) > \phi(U^1) > \cdots \phi(U^t) > \phi(U^{t+1}) > \cdots \geq \phi(\mathcal{G}).
\label{eqn:conv}
\end{equation}
Notice that $U^1, U^2, \cdots, U^t,\cdots$ are only a few of all the possible states. Since the update on $U^t$ happens only when a closer neighbor is found, the iteration leads function $\mathit{\phi(U^t)}$ to decrease monotonically. Meanwhile, the function value is lower-bounded by $\phi(\mathcal{G})$. It is therefore clear to see the iteration in S-Merge converges.

Both S-Merge and J-Merge are greedy optimization approaches. Similar to NN-Descent, S-Merge and J-Merge perform hill-climbing NN search~\cite{icai11:kiana} in a batch fashion. The greedy hill-climbing process could be trapped in local optima. While different from NN-Descent, S-Merge and J-Merge climb from the half-way up to the hill. Both S-Merge and J-Merge are effective on low dimensional data. As will be revealed later, their effectiveness drops as the data dimensionality increases.

The time complexity of NN-Descent is $O(d{\cdot}n^\rho)$, where $1.4 \leq \rho \leq 1.9$ according to our empirical study. In general, the cost of merging two graphs is lower than constructing the whole from scratch by NN-Descent. Given an n-scale dataset, we now have three ways to construct a \textit{k}-NN graph by NN-Descent. The first way builds the graph by NN-Descent directly. In a second way, the construction is decomposed into three steps. The first two steps construct the sub-\textit{k}-NN graphs for $S_1$ and $S_2$ by calling NN-Descent respectively, given ${\mid}S_1{\mid}={\mid}S_2{\mid}$. In the \textit{3}rd step, S-Merge is called to merge two subgraphs. The number of comparisons in S-Merge is the number of comparisons of the first way excluding out the number of comparisons that happens between samples from the same subset. Since samples from two subsets have been mixed up sufficiently in the initialization stage, the chance that one sample in a \textit{k}-NN list compares to a sample from the same subset is on the same level as the chance of being compared to a sample from another subset. Given ${\mid}S_1{\mid}={\mid}S_2{\mid}$, the number of cross-comparisons between two subsets takes up one-third of the total comparisons. Namely, we have

\begin{equation}
\begin{aligned}
P_s(n)=\frac{d{\cdot}n^{\rho}}{3}.
\end{aligned}
\label{eqn:tmpmerge}
\end{equation}

Since two subsets are already of high quality, the NN-Descent procedure converges faster than it runs on a completely raw set. The number of cross-comparisons between two subsets could be slightly smaller than $P_s(n)$. As a result, the time complexity upper bound of S-Merge is around one-third of NN-Descent.

In a third way, alternatively, J-Merge could be called to replace the \textit{2}nd and the \textit{3}rd steps in the second way. In this case, in addition to the number of cross-comparisons, the number of comparisons within the raw set should be counted. Since samples from two subsets have been mixed up sufficiently in the initialization stage, one sample in a \textit{k}-NN list holds an equal chance to compare to a sample from the same subset and a sample from another subset. As a result, the number of comparisons in J-Merge takes up  two-thirds of the total comparisons in NN-Descent. So we have the time complexity of J-Merge as

\begin{equation}
\begin{aligned}
P_j(n)=\frac{2{\cdot}d{\cdot}n^{\rho}}{3}.
\end{aligned}
\label{eqn:tmjmerge}
\end{equation}
Since subgraph $G_{1}$ is already in good shape, the NN-Descent iteration procedure converges faster. The number of comparisons required in J-Merge is therefore slightly smaller than $\frac{2{\cdot}n^{\rho}}{3}$ .

\begin{table}
\begin{center}
\caption{Time Complexity of S-Merge, J-Merge, and H-Merge. $\rho$ is usually in the range of $[1.4, 1.9]$. It varies as the scale of the data and the type of the data change.}
\label{tab:cmpx}
	\begin{tabular}{|l|c|c|c|c|} \hline
	Algorithm & S-Merge & J-Merge & H-Merge & NN-Descent \\ \hline\hline
	Time Comp. & $O(\frac{d{\cdot}n^{\rho}}{3})$ & $O(\frac{2{\cdot}d{\cdot}n^{\rho}}{3})$ & $O(\frac{5{\cdot}d{\cdot}n^{\rho}}{3})$ & $O(d{\cdot}n^{\rho})$ \\ \hline
	\end{tabular} \\
\end{center}
\end{table}

In S-Merge and J-Merge, the extra time has to be spent on merge sort at the final step. Since each \textit{k}-NN list is sorted, the merge sort is very efficient. The time complexity of this step is $O(n{\cdot}k)$. When both \textit{n} and \textit{d} are large, the time cost of this step is minor in contrast to the iteration procedure. Overall, the time complexity of S-Merge and J-Merge is around \textit{30\% }$\sim$\textit{70\%} of NN-Descent.

Before we discuss the time complexity of H-Merge, let's consider the time complexity of building a series of approximate \textit{k}-NN graphs with growing sizes by calling NN-Descent repetitively. Given the size of the graph is doubled each time, the scale of the problem is given as
\begin{equation}
	s=\cdots+\frac{n}{8}+\frac{n}{4}+\frac{n}{2}+n,
\end{equation}
where \textit{n} is the size of the dataset. As a result, the overall scale of the problem is $2{\cdot}n$. So the time complexity of building a hierarchy by the repetitive calling of NN-Descent is equivalent to building the whole graph twice by NN-Descent. Compared to this problem scale, H-Merge is more cost-saving since it performs J-Merge with an upper layer graph on each level. The scale of the problem is roughly two-thirds of NN-Descent of the same level as discussed before. As a consequence, the overall time complexity of H-Merge is roughly \textit{1.67} times over the complexity of calling NN-Descent to build a flat graph. The time complexity of S-Merge, J-Merge, and H-Merge are summarized in Tab.~\ref{tab:cmpx}.

\section{NN search over \textit{k}-NN Graph Hierarchy}
\label{sec:srch}
As demonstrated by HNSW graphs~\cite{pami18:yury}, the hierarchical graph structure is helpful for the NN search task. In this section, we are going to show that a similar hierarchical NN search mechanism could be undertaken with the graphs built by H-Merge.

As mentioned in the previous section, on each layer of the hierarchy, H-Merge produces an approximate \textit{k}-NN graph. It is possible to conduct the top-down NN search directly based on the hierarchy formed by a series of approximate \textit{k}-NN graphs. Similar to HNSW, the search starts from the top layer and performs hill-climbing on the graph until it reaches the closest sample on the layer. Thereafter the closest sample is taken as the starting point in the next layer search. The same search procedure is repeated on the second layer with the supplied seed. This process continues until it finds the closest sample on the bottom layer.

Although the above search scheme is feasible, the comparison redundancy happens due to the occlusions in the \textit{k}-NN neighborhood~\cite{cvpr16:ben,dpg:wenli,pami18:yury}. Following the practice in~\cite{cvpr16:ben,dpg:wenli, pami18:yury}, the graph diversification is applied to the approximate \textit{k}-NN graph of each layer. This leads to a sparse NN graph and therefore reduces the potential comparison redundancy. In this paper, the graph diversification (GD) operation scheme used in HNSW~\cite{cvpr16:ben,pami18:yury} is adopted. However different from~\cite{pami18:yury}, the graph diversification is performed on the already built approximate \textit{k}-NN graphs as a post-processing step. In contrast, HNSW applies the diversification on an incrementally diversified graph~\cite{pami18:yury}.

\begin{figure}
	\begin{center}
		\includegraphics[width=0.3\linewidth]{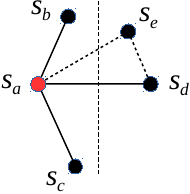}
	\end{center}
	\caption{The illustration of the heuristic strategy used to select the graph neighbors in each layer of the hierarchical structure. An edge from $s_a$ to $s_d$ occludes an edge from $s_a$ to $s_e$ since $s_e$ is closer to $s_d$ than $s_a$.}
	\label{fig:delaunay}
\end{figure}

As shown in Alg.~\ref{alg:gd}, GD examines the neighborhood of each \textit{k}-NN list. Given sample $s_a$, the nearest neighbor of $s_a$ is kept by default. The rest neighbors are treated as the candidates to be examined. The candidates are sorted in ascending order and are examined one by one. A candidate is kept if its distance to sample $s_a$ is smaller than its distances to all the already kept samples. As illustrated in Fig.~\ref{fig:delaunay}, sample $s_e$ is removed from the NN list of sample $s_a$, since its distance to $s_d$ is smaller than it is to $s_a$. The reverse \textit{k}-NN list of sample $s_a$ is also diversified based on the same rule and merged with the diversified \textit{k}-NN list. The \textit{k}-NN graph of each layer is undergone this diversification operation. The hierarchy after graph diversification is ready to support efficient NN search.

\begin{figure}
	\begin{center}
		\includegraphics[width=0.55\linewidth]{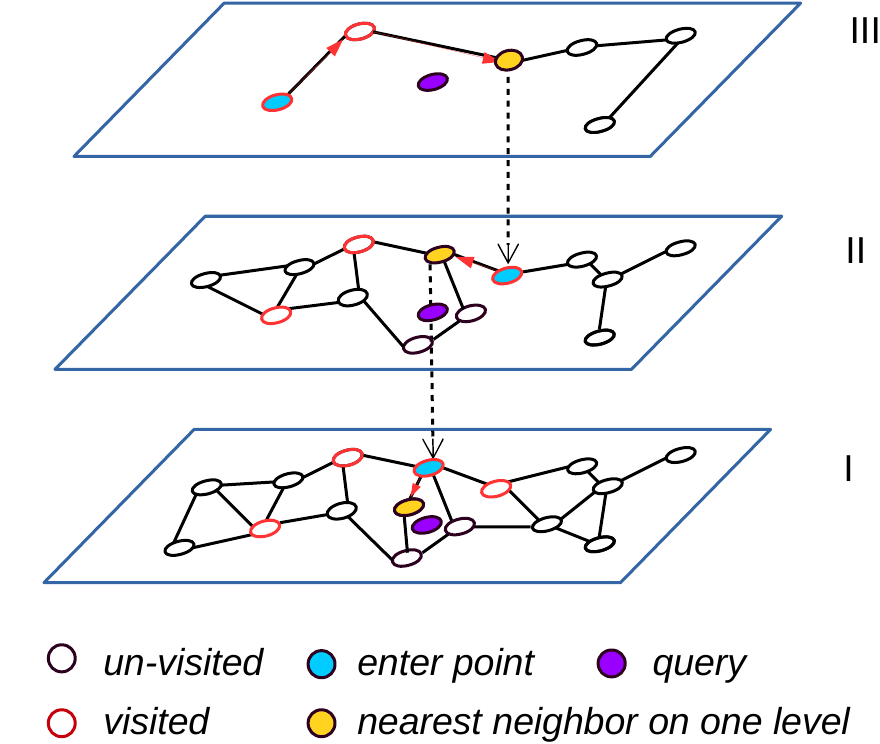}
	\end{center}
	\caption{The illustration of top-down NN search on a three-layer graph hierarchy.}
	\label{fig:hier}
\end{figure}

The NN search on the hierarchy largely follows the procedure of HNSW~\cite{pami18:yury}. It is divided into two stages. At the first stage, a coarse greedy search is conducted on the non-bottom layers. The search starts from the top layer where we have the smallest graph. It starts from one random sample on the layer and explores the neighbors of a visited vertex. The closest neighbor in the expanded neighborhood is treated as the candidate to explore in the next round. The search stops when no closer neighbor to the query is found on this layer. The closest neighbor found at the upper layer is treated as the starting point of the search on the next layer. This process repeats until it reaches the last non-bottom layer. In the second stage, the discovered closest sample on the first phase is taken as the starting point for the bottom-layer search. Different from the non-bottom layer search, all the samples maintained in a top-ranked list are expanded during the search. The search visits the top-ranked list in a best-first fashion and terminates when no new sample in the rank list to be expanded. The search procedure on a three-layer hierarchy is illustrated in Fig.~\ref{fig:hier}.

Similar to HNSW, the links (edges) kept in the upper layers connect vertices relatively far away from each other. Compared to the NN search on a single flat graph, the search moves faster when it is undertaken on these coarser graphs. Therefore the NN search on the non-bottom layers is expected to supply candidates to the bottom layer search in a more efficient way than search on a flat graph.

\begin{algorithm}
  \KwData{ $nhood$: neighborhood of sample $s_a$}
  \KwResult{candidate set $Q$ selected by heuristic}
  $Q \leftarrow nhood[1]$ \;
  \For{each $s_i \in nhood[2\cdots{k}]$}
  {
  		$flag= true$ \;
  		\For{each $c \in Q$}
  		{
  			\If{$m(s_i, c) < m(s_i, s_a)$}
  			{	$flag= false$\;
  				break;
  			}
  		}
  		\If{$flag$}
  			{	
  			 $Q \leftarrow Q \cup s_i$\;
  			}
  }
  \Return $Q$
  \caption{\textit{k}-NN Graph Diversification}
 \label{alg:gd} 
\end{algorithm}

\begin{table*}[!h]
\scriptsize{
\begin{center}
\caption{Overview of Datasets.}
\
\footnotesize{
\begin{tabular}{|l|lrr|rr|ll|}
\hline
Dataset & n &d & Qry & Exh. Srch (s) & LID~\cite{nips05:Levina} & m($\cdot$,$\cdot$)&Type \\
\hline \hline

RAND100K&$1\times10^5$&$2\sim100$&- &-&-&$\textit{l}_2$&synthetic\\ 
RAND100K&$1\times10^5$&$2\sim100$&- &-&-&$\textit{l}_1$&synthetic\\
SIFT100K~\cite{JDS11}& $1{\times}10^5$& 128 & - & - & - & $\textit{l}_2$ & SIFT~\cite{ijcv04:sift} \\
MNIST~\cite{ann-benchmarks}& $7{\times}10^4$ &784 & - & - & - & $\textit{l}_2$ & Image \\
NYTimes~\cite{ann-benchmarks}& $3{\times}10^5$ &256 & - & - & - & \textit{Cosine} & Text \\
Kosarak~\cite{ann-benchmarks}& $7.4{\times}10^4$ &27,983 & - & - & - & \textit{Jaccard} & Itemset \\ \hline \hline

RAND10M8D & $1\times10^7$ & 8 &1,000&78.8& 6.5&$\textit{l}_2$&synthetic  \\ 
RAND10M32D & $1\times10^7$ & 32 &1,000&26.1&19.4 &$\textit{l}_2$&synthetic  \\ 
RAND1M100D & $1\times10^6$ & 100 &1,000&92.0&48.9 &$\textit{l}_2$&synthetic \\ 
NUSWide & $2.6\times10^5$ & 500 &1,000&958.1& 38.2&$\kappa^2$&BoVW ~\cite{iccv03:sivic} \\ 

SIFT1M~\cite{JDS11} & $1\times10^6$  & 128 & 10,000 &1184.9& 16.3&$\textit{l}_2$&SIFT~\cite{ijcv04:sift} \\ 

SIFT10M~\cite{JDS11} & $1\times10^7$  & 128 & 10,000 &11441.0& 16.3&$\textit{l}_2$&SIFT~\cite{ijcv04:sift} \\ 

GIST1M~\cite{civr09:jegou}& $1\times10^6$ & 960 &1,000&919.1&38.1 &$\textit{l}_2$&GIST~\cite{civr09:jegou} \\ 
GloVe1M~\cite{glove14} & $1.2\times10^6$ & 100 &1,000&109.3 & 39.5 &\textit{Cosine}&Text\\ 

YFCC1M~\cite{yfcc100m} & $1\times10^6$ &128 &10,000&1091.6 & 23.4 &$\textit{l}_2$&Deep Feat.\\ 

Kosarak~\cite{ann-benchmarks}& $7.4{\times}10^4$ &27,983 & 500 & 66.5 & 47.3 & \textit{Jaccard} & Itemset \\ \hline

\end{tabular}
}
\label{tab:datasets}
\vspace{-0.2in}
\end{center}
}
\end{table*}

\section{Experiments}
\label{sec:exp}
In this section, the performance of S-Merge and J-Merge is studied on the \textit{k}-NN graph merge task. Each of the datasets is divided into two subsets. The sub-\textit{k}-NN graphs are constructed in advance by NN-Descent. S-Merge and J-Merge are called respectively to fulfill the merge in two different scenarios. Since there is no graph merge algorithm in the literature, their performance is compared to NN-Descent, NSW \cite{infosys13:yury} and RLB \cite{jmlr09}, each of which is adopted to construct the graph for the whole set directly. While H-Merge is evaluated when it is used as a hierarchical approximate \textit{k}-NN graph construction approach. Meanwhile, the performance of H-Merge is also studied when the hierarchical structure is used to support fast NN search. 

As revealed in~\cite{weidong,dpg:wenli}, the complexity of approximate \textit{k}-NN graph construction is largely related to the intrinsic data dimension. The intrinsic data dimension of real-world data varies considerably from one data type to another. Different from real-world data, the intrinsic dimension of synthetic data increases steadily as the data dimension increases. As a result, the performance trend of \textit{k}-NN merge approach on synthetic data is more observable compared to the real-world data. For this reason, the experiments of \textit{k}-NN graph merge are conducted on a series of synthetic datasets. The data dimension varies from \textit{2} to \textit{100}, which is in line with the convention in~\cite{weidong}. Values in each dimension are independently drawn from the range $[0,~1)$ under the uniform distribution. Moreover, another four real-world datasets are further adopted to demonstrate the performance of S-Merge and J-Merge on different data types. Datasets MNIST and Kosarak are composed of sparse vectors. The densities of these two datasets are \textit{0.198\%} and \textit{19.270\%} respectively.

For the NN search task, the performance is reported on both synthetic random data and data from the real-world. Besides three large-scale synthetic datasets, seven real-world datasets are adopted in the evaluation. The brief information about all the datasets is summarized in Tab.~\ref{tab:datasets}. Among them, seven datasets are derived from real-world images, deep features, or text data. Four datasets, namely GIST1M, GloVe1M, NUSWide, and Rand1M that are marked as most challenging datasets in~\cite{dpg:wenli}, are incorporated. For each of the datasets, another set of queries of the same data type are prepared. Different metrics such as $\textit{l}_2$, \textit{Cosine}, $\kappa^2$, and \textit{Jaccard} are adopted in accordance with the data type of each set.

On the NN search task, the performance of the proposed search approach is studied in comparison to the representative approaches of different categories. Namely, they are graph-based approaches such as DPG~\cite{dpg:wenli}, SPTAG\footnote{In favor of search quality, SPTAG-BKT is used in the experiment.}~\cite{sptag} and HNSW~\cite{pami18:yury}. SRS~\cite{srs14} is considered as the representative locality-sensitive hash approach. Product quantizer (PQ)~\cite{JDS11} is considered as the representative quantization based approach in the comparison. FLANN~\cite{pami14:flann} and Annoy~\cite{annoy} are selected as the representative tree partitioning approaches, both of which are popular NN search libraries in the literature.

\subsection{Evaluation Protocol}
For \textit{k}-NN graph construction, the top-\textit{1} (\textit{recall@1}) and top-\textit{10} (\textit{recall@10}) recalls on each dataset are studied under $\textit{l}_1$ and $\textit{l}_2$ metrics respectively. Given function $\mathit{R(i,k)}$ returns the number of truth-positive neighbors at top-\textit{k} NN list of sample $i$, the recall at top-$k$ on the whole set is given as 
\begin{equation}
	recall@k=\frac{\sum_{i=1}^n{R(i,k)}}{n{\times}k}.
\label{eval:recall}
\end{equation}

Besides \textit{k}-NN graph quality, the construction cost is also studied by measuring the scanning rate~\cite{weidong} of each approach. Given $C$ is the total number of distance computations in the construction, the scanning rate is defined as
\begin{equation}
c=\frac{C}{n{\times}(n-1)/2}.
\label{eval:scan}
\end{equation}

The search quality is measured by the top-\textit{1} recall for the first nearest neighbor. This is in line with the evaluation convention in the literature. Moreover, in order to allow the readers to know how efficient that one NN search approach performs, the time cost for brute-force NN search on each dataset is also shown on the \textit{5}th column of Tab.~\ref{tab:datasets}.

All the codes of different approaches considered in this study are compiled by g++ \textit{5.4}. The multi-threads, SIMD, and pre-fetching instructions are enabled as long as they are supported in the source codes for NN search task. While for \textit{k}-NN graph merge task, OpenMP~\cite{openmp} is adopted. This is to illustrate our merge algorithms are parallelizable on the CPU threads level as NN-Descent. All the experiments are pulled out on a workstation with \emph{24} cores of \textit{2.4}GHz CPUs and \textit{64}G memory setup.

\subsection{Performance of S-Merge and J-Merge}
In this section, the performance of two proposed merge algorithms is studied in comparison to NN-Descent~\cite{weidong}, which is recognized as the state-of-the-art approximate \textit{k}-NN graph construction algorithm. The evaluation is conducted on six synthetic datasets. While the size of the datasets is fixed to \textit{100}K, which is in line with~\cite{weidong}. The performance is reported when the merge algorithms are operated under $l_1$ and $l_2$ metrics respectively. The graph quality achieved by S-Merge, J-Merge, and NN-Descent is largely controlled by \textit{k}, which is exactly the length of the NN list. In our experiment, \textit{k} is set to the value where the algorithm makes a balance between efficiency and quality.

Each dataset is divided into two subsets. For S-Merge, NN-Descent is called to build two sub-\textit{k}-NN graphs respectively for two subsets. For J-Merge, NN-Descent is called to build the \textit{k}-NN graph for one of the subsets, while leaving another as the raw set. It is possible to use approach~\cite{olg:wlzhao} to produce the subgraph, which actually achieves better quality. However, NN-Descent is preferred in the test as both S-Merge and J-Merge can be viewed as the extensions over NN-Descent. It is easy to see the efficiency and quality that S-Merge and J-Merge achieve over NN-Descent. 

In S-Merge and J-Merge, the number of samples in each NN list that is cut out for the final merge could range from \textit{0} to \textit{k}. Given the number of samples reserved for mixture-up is $k_1$, the ratio that regularizes this division is defined as $r=\frac{k_1}{k}$. This factor impacts the performance of both S-Merge and J-Merge. In the following, an ablation analysis is made to see how this factor impacts the performance of the graph merge in two scenarios.

\begin{figure}
\begin{center}
 \subfigure[Symmetric Merge]
 {\includegraphics[width=0.48\linewidth]{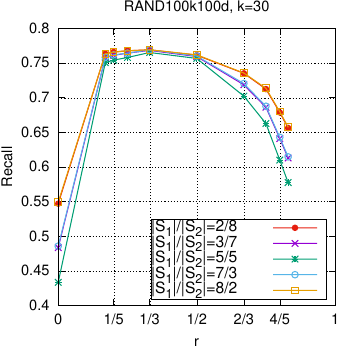}}
 \hspace{0.01in}
	\subfigure[Joint Merge]
 {\includegraphics[width=0.48\linewidth]{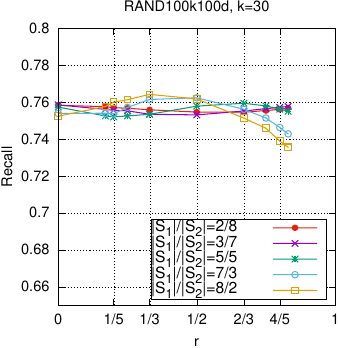}}
 
\caption{The fluctuation trend of graph quality when varying parameter $r$ in two merge strategies on RAND100K100D dataset. Parameter \textit{k} is fixed to \textit{30}. Figure (a) and Figure (b) show the variation of recall@10 when parameter $r$ changes from \textit{0} to \textit{4/5} for S-Merge and J-Merge respectively. In both cases, the size ratio of between $S_1$ and $S_2$ is varied from $2/8$ to $8/2$. To ensure the statistical significance, the digits are reported by averaging \textit{20} rounds of results for each parameter setting.}
\label{fig:parameter}
\end{center}
\end{figure}

\begin{table*}
\scriptsize{
\begin{center}

\caption{Construction scanning rate (\textit{c}) and wall-time (\textit{tm}) in seconds (s) of S-Merge and J-Merge in comparison to NN-Descent, NN-Descent*, NSW, and RLB on six synthetic and four real-world datasets. The dimension (\textit{d}) of these synthetic datasets ranges from 2 to 100. For S-Merge and J-Merge, ${\mid}S_1{\mid}={\mid}S_2{\mid}$. All the approaches are tested on a single core.}
\footnotesize{
\begin{tabular}{|r|r|r||rr||rr||rr||rr||rr||rr|}
\hline
\multicolumn{3}{|c||}{Dataset} & \multicolumn{2}{c||}{NN-Desc.} & \multicolumn{2}{c||}{NN-Desc.*} & \multicolumn{2}{c||}{NSW} & \multicolumn{2}{c||}{S-Merge} & \multicolumn{2}{c||}{J-Merge} &\multicolumn{2}{c|}{RLB} \\
\hline
m($\cdot$) & d & \textit{k} & c & tm & c & tm & c & tm & c & tm & c & tm & c & tm\\
\hline \hline
\multirow{5}{*}{$l_{1}$} 
& $2$  &10 & 0.009 & \textbf{2.611} & 0.003 & 6.200 & 0.006 & 3.031 & 0.006 & 3.996 &0.006 & 2.755 & \textbf{0.001} & 7.734\\ 
~ & $5$ &10 & 0.010 & 2.995 & 0.003 & 6.734 & 0.006 & \textbf{2.457} & 0.004 & 2.988 &0.007 & 3.169 & \textbf{0.003} & 19.334\\ 
~ & $10$&20 & 0.038 & 8.656 & 0.006 & 19.111 & 0.019 & 6.247 & 0.012 & \textbf{5.570} & 0.021 & 7.796 & \textbf{0.004} & 32.862\\ 
~& $20$ &20 & 0.048 &12.836 & 0.014 & 19.408 & 0.028 & 9.216 & 0.014 & \textbf{7.827} & 0.028 & 10.748 & \textbf{0.009} & 114.113\\
~& $50$ &30 & 0.120 &42.124 & 0.052 & 83.090 & 0.075 & 34.268 & \textbf{0.034} & \textbf{18.444} &0.071 & 30.786 & 0.066 & 1475.360\\ 
~& $100$&40 & 0.208 &120.066 & 0.087 & 185.087 & 0.127 & 99.205 & \textbf{0.059} & \textbf{42.394} &0.121 & 79.437 & 0.141 & 5081.260\\ \hline \hline

\multirow{5}{*}{$l_{2}$} 
& $2$  &10 & 0.008 & \textbf{2.673} & 0.003 & 6.004 & 0.006 & 2.928 & 0.006 & 4.061 &0.006 &2.763 & \textbf{0.001} & 7.183\\
~ & $5$ &10 & 0.010 & 3.122 & 0.003 & 6.052 & 0.005 & \textbf{2.221} & 0.004 & 2.979 &0.007 &3.090 & \textbf{0.002} & 18.374\\
~ & $10$&20 & 0.036 & 7.559 & 0.006 & 10.126 & 0.019 & 5.894 & 0.013 & \textbf{5.759} &0.022 &7.787 & \textbf{0.003} & 31.502\\ 
~& $20$ &20 & 0.051 & 13.264 & 0.014 & 19.671 & 0.028 & 9.186 & 0.015 & \textbf{8.093} &0.030 &11.092 & \textbf{0.007} & 116.916\\
~& $50$ &30 & 0.129 & 44.398 & 0.055 & 88.129 & 0.079 & 38.992 & \textbf{0.038} & \textbf{19.859} &0.077 &32.378 & 0.067 & 1469.280\\ 
~& $100$&40 & 0.216 & 126.386 & 0.079 & 175.916 & 0.150 & 139.409 & \textbf{0.064} & \textbf{45.972} &0.126 &83.926 & 0.141 & 5093.500\\ \hline

\end{tabular}

\begin{tabular}{|rr|r||rr||rr||rr||rr||rr||rr|} \hline\hline

 & & \textit{k} & c & tm & c & tm & c & tm & c & tm & c & tm & c & tm\\
\hline \hline
\multicolumn{2}{|l|}{SIFT100K} & 20 & 0.056  & 45.019 & 0.022 & 66.078 & 0.029 & 26.551 & 0.019 & \textbf{19.426} & 0.035 & 31.074 & \textbf{0.004} & 53.918\\ 
\multicolumn{2}{|l|}{MNIST} & 40 &0.191 & 368.498 & 0.017 & \textbf{71.542} & 0.242 & 254.647 & 0.067 & 136.504 & 0.115 & 227.231 & \textbf{0.012} & 379.126\\ 
\multicolumn{2}{|l|}{NYTimes} & 40 & 0.065 & 836.950 & 0.026 & 1020.780 & 0.033 & 665.268 & 0.021 & \textbf{313.516} & 0.038 & 528.336 & \textbf{0.012} & 4091.490\\ 
\multicolumn{2}{|l|}{Kosarak} & 40 & 0.243 & 963.505 & 0.154 & 1140.600 & - & - & \textbf{0.088} & \textbf{343.802} & 0.150 & 557.900 & - & - \\ \hline
\end{tabular}

}
\label{tab:merge_time}

\vspace{-0.2in}
\end{center}

}
\end{table*}

\subsubsection{Ablation Analysis}
\label{sec:expaba}
The ablation analysis about parameter \textit{r} is conducted on ``RAND100K100D'' with $l_2$ distance measure. \textit{k} is set to \textit{30}. As mentioned before, NN-Descent is called to produce subgraphs. Then S-Merge and J-Merge are called to merge the approximate sub-\textit{k}-NN graphs. The size ratio between two subsets, namely $\frac{{\mid}S_1{\mid}}{{\mid}S_2{\mid}}$ is varied from $\frac{2}{8}$ to $\frac{8}{2}$. So that both the balanced and unbalanced cases of graph merge are studied. Under each size ratio, different settings for parameter \textit{r} are tested.

Fig.~\ref{fig:parameter}(a) and Fig.~\ref{fig:parameter}(b) show the curves of the graph quality variation under each size ratio when \textit{r} varies from \textit{0} to $\frac{4}{5}$ for S-Merge and J-Merge respectively. As shown in Fig.~\ref{fig:parameter}(a), the graph quality from S-Merge is maintained on the high level only when \textit{r} is in the range $[\frac{1}{5}, \frac{1}{2}]$. Within this range, it allows the samples from two sides to be exposed to the cross-comparison sufficiently. In contrast, J-Merge is less sensitive to \textit{r}. As shown in Fig.~\ref{fig:parameter}(b), in the case ${\mid}S_1{\mid} \leq {\mid}S_2{\mid}$, the graph quality varies a little even when no samples from $S_2$ are joined into the \textit{k}-NN list of subgraph $G_1$. This is because samples from $S_1$ and $S_2$ are already present in the initial graph $G_2$. It already guarantees that the samples from both sides are sufficiently exposed to cross-comparison. In the case ${\mid}S_1{\mid} > {\mid}S_2{\mid}$, the initial graph $G_2$ is dominated by samples from $S_1$. The raw samples have to be joined into the \textit{k}-NN list of $G_1$ to be better exposed to the cross-comparison. As shown in Fig.~\ref{fig:parameter}(b), a valid range for \textit{r} in J-Merge is $[\frac{1}{3}, \frac{1}{2}]$. In the meantime, the higher chance of being exposed to cross-comparison induces more number of distance computations. This in turn leads to the higher computation cost as well as the higher graph quality. In practice, a valid setting for \textit{r} is $\frac{1}{2}$ in both merge algorithms as we want to maintain similar graph quality as NN-Descent.

\begin{figure}
\begin{center}
 \subfigure[Recall@1 with $l_{1}$-norm]
 {\includegraphics[width=0.48\linewidth]{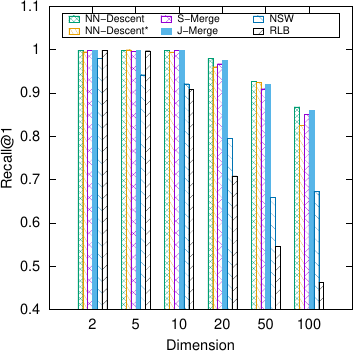}}
 \hspace{0.01in}
	\subfigure[Recall@10 with $l_{1}$-norm]
 {\includegraphics[width=0.48\linewidth]{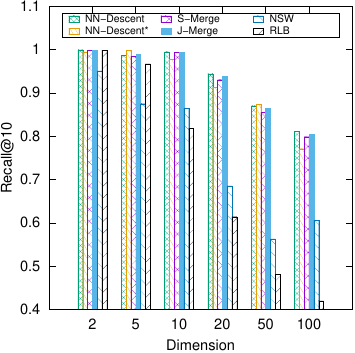}}
 \subfigure[Recall@1 with $l_{2}$-norm]
 {\includegraphics[width=0.48\linewidth]{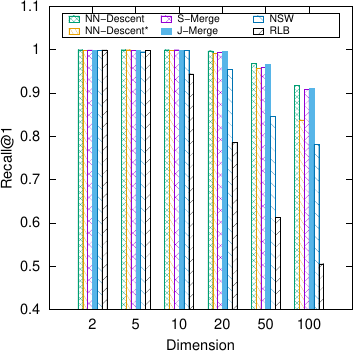}}
 \hspace{0.01in}
	\subfigure[Recall@10 with $l_{2}$-norm]
 {\includegraphics[width=0.48\linewidth]{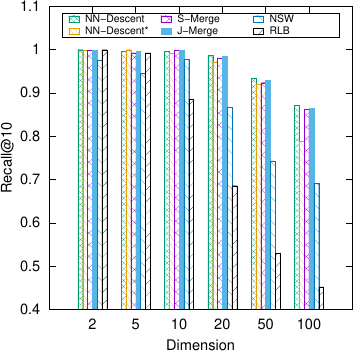}} 
 \subfigure[Recall@1]
 {\includegraphics[width=0.48\linewidth]{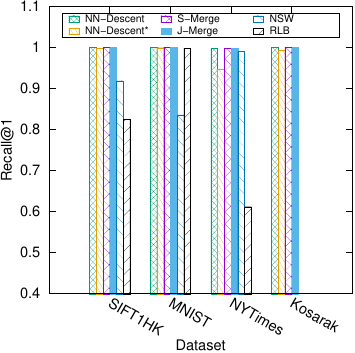}}
 \hspace{0.01in}
 \subfigure[Recall@10]
 {\includegraphics[width=0.48\linewidth]{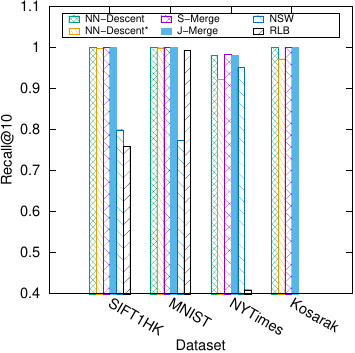}} 
 \caption{The quality of approximate \textit{k}-NN graphs produced by S-Merge and J-Merge, in comparison to NN-Descent, NSW and RLB on six synthetic datasets and four real-world datasets. On the synthetic datasets, both $\textit{l}_1$-norm and $\textit{l}_2$-norm are tested. On all the runs for S-Merge and J-Merge, ${\mid}S_1{\mid}={\mid}S_2{\mid}$.}
\label{fig:merge_performance}
\end{center}
\end{figure}

\subsubsection{Performance on approximate \textit{k}-NN graph Merge}
In this section, the effectiveness, as well as the efficiency of the two merge algorithms, are studied on both the synthetic and real-world datasets. For convenience, the performance is evaluated in the scenario of merging two subsets with equal size. In the test for S-Merge, two sub approximate \textit{k}-NN graphs are prepared by NN-Descent in advance. S-Merge is called to merge the two. In the test for J-Merge, one sub approximate \textit{k}-NN graph is prepared by NN-Descent. J-Merge is called to merge the left raw set into the subgraph. Their performance is compared to incremental NN-Descent (given as NN-Descent*), online \textit{k}-NN graph construction approach NSW~\cite{infosys13:yury}, and divide-and-conquer \textit{k}-NN graph construction approach RLB~\cite{jmlr09}. For NN-Descent*, the first half \textit{k}-NN graph is constructed by NN-Descent. The rest of the samples are inserted into the \textit{k}-NN graph one by one by the NN search procedure provided in the same package~\cite{weidong}. NN-Descent* and J-Merge functionally overlap with each other. Since no mechanism is designed in NSW to merge two existing graphs, samples from one sub-graph are inserted into another by NSW algorithm one by one. In another word, both joint merge and symmetric merge are treated as joint merge in NSW. The parameter \textit{R} of the RLB code\footnote{Codes are collected from https://jiechenjiechen.github.io.} is tuned (from 0.07 to 0.3) with respect to different data dimensions. Parameter \textit{k} in RLB is fixed to 10, the setting of which is a trade-off between graph quality and time cost. The performance from NN-Descent is still presented when it is called to produce an approximate \textit{k}-NN graph directly for the whole set. This is to show how much cost the graph-merging algorithm could save compared to building the graph from scratch.

For different datasets, the parameter \textit{k} for NN-Descent, NN-Descent*, NSW and our merge algorithms varies. It is largely close to the data dimension. While \textit{k} is shared the same across different approaches on the same dataset. The first evaluation is conducted on six synthetic datasets and four real-world datasets. Top-\textit{1} and top-\textit{10} recalls for all five approaches are shown in Fig.~\ref{fig:merge_performance}. Correspondingly, the parameter setting for \textit{k}, the scanning rate, and the wall-time for all the runs are presented in Tab.~\ref{tab:merge_time}.

As shown in the figure, the graph quality achieved by the two merge algorithms is largely similar to NN-Descent on both synthetic and real-world datasets. On the synthetic datasets, the graph quality that is produced by NN-Descent is slightly higher when the data dimension is in the range of $[10, 100]$. However, the performance difference between them is within \textit{3\%}. 
On the other hand, the scanning rates of S-Merge and J-Merge are below the theoretical baselines that we discussed in Section~3.4. NN-Descent shows the highest scanning rate. The scanning rates from J-Merge are always higher than that of S-Merge across all the runs. However, according to our observation, S-Merge actually shows slightly higher scanning rates than J-Merge if the number of comparisons spent on building subgraphs is also considered. In the meantime, J-Merge shows slightly better performance than S-Merge. This trend is subtle but consistent. This does indicate it is more cost-effective by calling J-Merge than calling NN-Descent followed by S-Merge to merge a raw set. When a raw set is joined into a built graph, the raw samples are moving along a structured graph, which is easier for them to find true close neighbors. Compared to J-Merge, NN-Descent* shows lower scanning rate while considerably higher time costs. This trend is consistent except for MNIST, which is in low intrinsic dimension and small scale. In NN-Descent*, in addition to distance computation, extra cost is induced to maintain a dynamic reverse \textit{k}-NN graph. Meanwhile, J-Merge outperforms NN-Descent* particularly on high dimensional datasets. The former is more cost-effective as it performs the hill-climbing in batch.

As shown in Fig.~\ref{fig:merge_performance}, both NSW and RLB show considerably poorer graph quality than the other approaches across all the datasets. While as shown in Tab.~\ref{tab:merge_time}, the computation cost of NSW is close to J-Merge when the data dimension is low. As the data dimension increases, its computation cost becomes much higher than J-Merge. It is clear to see that fast online \textit{k}-NN graph construction is not the appropriate solution for the graph merge problem. The major difference between J-Merge and NSW is that J-Merge performs ``hill-climbing'' batchfully. RLB only performs relatively well on high-dimensional data with low intrinsic dimension such as MNIST. NSW and RLB perform much poorer than NN-Descent and our merge algorithms. Comparing Fig.~\ref{fig:merge_performance}(a)-(b) with Fig.~\ref{fig:merge_performance}(a)-(b), the performance varies across different distance metrics for the data on the same dimension. The data are distributed differently under different metric spaces. \textit{k}-NN graph construction for data in $\textit{l}_1$-space is harder than that of $\textit{l}_2$-space, where the data samples are easier to reach to each other via neighborhood links.

\begin{figure}
\begin{center}
 \subfigure[SIFT1M]
 {\includegraphics[width=0.45\linewidth]{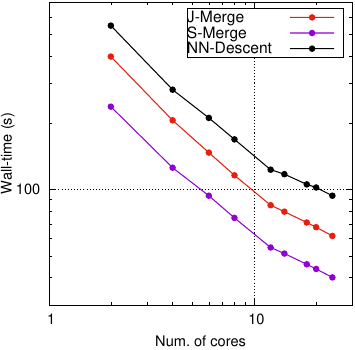}} 
 \hspace{0.15in}
 \subfigure[GIST1M]
 {\includegraphics[width=0.46\linewidth]{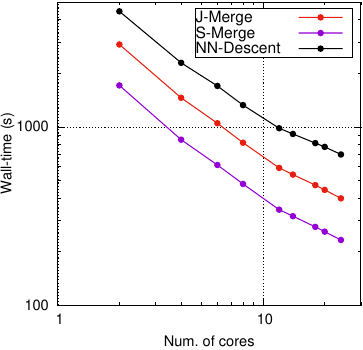}} 
 \caption{The efficiency of S-Merge and J-Merge in comparison to NN-Descent when they run on multiple threads. The experiment is conducted by increasing the number of cores incrementally from \textit{2} to \textit{24}. \textit{k} is fixed to \textit{30} on both datasets.}
\label{fig:paral}
\end{center}
\end{figure}

As seen from the performance reported on the real-world datasets (bottom table in Tab.~\ref{tab:merge_time}), most of the approaches show considerably higher scanning rates on sparse datasets MNIST and Kosarak than that of SIFT100K and NYTimes. This does indicate the greedy graph-based approaches are not the best option for sparse itemsets~\cite{icde16:boutet}.

In terms of running time, the costs from S-Merge and J-Merge are close to that of NN-Descent when the data dimension is low. As shown in Tab.~\ref{tab:merge_time}, the time costs from S-Merge and J-Merge are even slightly higher than NN-Descent for datasets of \textit{2}-dimension and \textit{5}-dimension. In these cases, the time costs spent on distance computation are close to the cost in the merge sort step, which causes extra computation overhead. Nevertheless, as the data dimension grows, the distance computation dominates the overall cost, while the time cost for merge sort becomes minor. As the data dimension is higher than \textit{20}, the time costs from S-Merge and J-Merge are around one-third and two-thirds of NN-Descent, which meets well with the analysis presented in Section~3.4.

\begin{table*}[t]
\scriptsize{
\begin{center}
\caption{Performance on \textit{k}-NN graph Construction for \textit{10} large-scale datasets. All the approaches run on \textit{24} cores. HNSW and SPTAG are not designed for \textit{k}-NN graph construction. The \textit{k}-NN graph quality from them is listed for reference only.}
\begin{tabular}{|l|rl||rl||rl||rl||rl|}
\hline
\multicolumn{1}{|c|}{\multirow{2}{*}{Dataset}} & \multicolumn{2}{c||}{RAND10M8D} & \multicolumn{2}{c||}{RAND10M32D} & \multicolumn{2}{c||}{RAND1M100D} & \multicolumn{2}{c||}{NUSWide} & \multicolumn{2}{c|}{SIFT1M} \\ 
\multicolumn{1}{|c|}{}             & Time (s)   & Recall@10    & Time (s)      & Recall@10    & Time (s) & Recall@10    & Time (s)      & Recall@10    & Time (s)     & Recall@10   \\ \hline \hline
H-Merge     & 597.07     & \textbf{1.000}     & 135.74     & 0.985     & 288.88     & 0.532  & 1372.52 & \textbf{0.869}   & 251.45    & \textbf{0.998}    \\ 
HNSW        & 205.36     & -       & 394.74     & 0.740     & 580.38     & 0.254   & 190.37 & 0.151 & 187.23    & 0.532    \\ 
NN-Descent    & \textbf{132.84}     & 0.998     & \textbf{56.65}     & \textbf{0.987}     & \textbf{154.38}     & \textbf{0.544}  & \textbf{757.68} &  0.849  & \textbf{136.07}    & 0.997    \\ 

SPTAG     & 6132.00     & 0.294 &2123.00     & 0.613     & 4555.00     & 0.479 & - & -    & 2490.00 & 0.478 \\

\hline \hline

\multicolumn{1}{|c|}{\multirow{2}{*}{Dataset}} & \multicolumn{2}{c||}{YFCC} & \multicolumn{2}{c||}{GIST1M} & \multicolumn{2}{c||}{GloVe1M} & \multicolumn{2}{c||}{Kosarak} & \multicolumn{2}{c|}{SIFT10M} \\ 
\multicolumn{1}{|c|}{}             & Time (s)   & Recall@10    & Time (s)      & Recall@10    & Time (s)      & Recall@10    & Time (s)      & Recall@10    & Time (s)     & Recall@10   \\ \hline \hline 

H-Merge                     & 264.90     & \textbf{0.980} & 1345.75     & \textbf{0.969}     & 369.37    & \textbf{0.921}    & 117.724 & 0.967    & 3549.86    &\textbf{0.995}    \\

HNSW  & 252.48    & 0.600    & \textbf{591.28}    & 0.252     & 304.82    & 0.462     & -     & -   & 2458.55    & 0.588    \\

NN-Descent                     & \textbf{133.91}     & 0.975     & 990.98    & 0.966     & \textbf{155.70}    & 0.913    & \textbf{76.677} & \textbf{1.000}     & \textbf{1620.41}	    & 0.993    \\

SPTAG                     & 3338.00     & 0.605 & 17352.00     & 0.224     & 3590.00   &0.293    & - &-     & 35150.00   &0.604    \\

 \hline
\end{tabular}
\label{tab:graph_recall}
\end{center}
}
\end{table*}

The efficiency of S-Merge and J-Merge is further studied on the multi-threads context. Similar to NN-Descent, both S-Merge and J-Merge are parallelizable on the CPU threads level. The experiment is pulled out on two large-scale datasets SIFT1M and GIST1M. Each dataset is divided into two subsets with equal size. Following the same pipeline as before, S-Merge and J-Merge run on multiple cores to merge the subgraphs. NN-Descent is treated as the comparison baseline.

Fig.~\ref{fig:paral} shows the wall-time of three algorithms as the number of cores increases incrementally.
As shown in the figure, all algorithms show steady speedup when the number of cores increases. The speedup becomes less significant when the number of cores reaches \textit{12}.

S-Merge and J-Merge are essentially the extensions over NN-Descent. Both of them maintain similar efficiency and effectiveness of NN-Descent. Although their overall costs and performance are similar, they play different roles in the approximate \textit{k}-NN graph construction. S-Merge is suitable for parallel or distributed approximate \textit{k}-NN graph construction. J-Merge allows an approximate \textit{k}-NN graph to be incrementally built. While the original NN-Descent is suitable for building a graph for a fixed dataset or build subgraphs for S-Merge and J-Merge use.

\begin{figure*}[t]
\begin{center}
	\subfigure[RAND10M8D]
	{\includegraphics[width=0.203\linewidth]{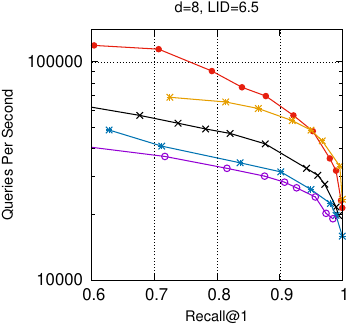}} 
	\hspace{0.01in} 
	\subfigure[RAND1M32D]
	{\includegraphics[width=0.188\linewidth]{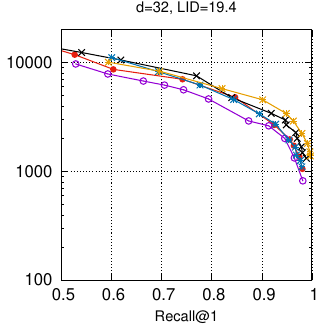}} 
	\hspace{0.01in} 
	\subfigure[RAND1M100D]
	{\includegraphics[width=0.185\linewidth]{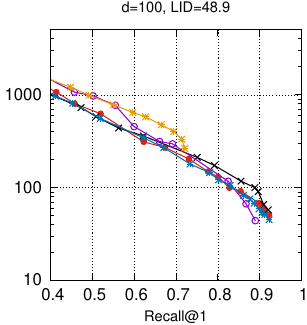}}
	\hspace{0.01in}
	\subfigure[NUSWide]
	{\includegraphics[width=0.177\linewidth]{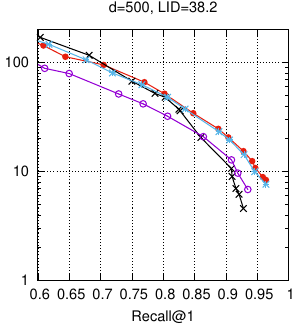}} 
	\hspace{0.01in} 
	\subfigure[SIFT1M]
	{\includegraphics[width=0.188\linewidth]{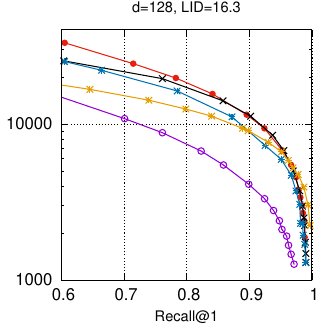}}
	\subfigure[YFCC]
	{\includegraphics[width=0.203\linewidth]{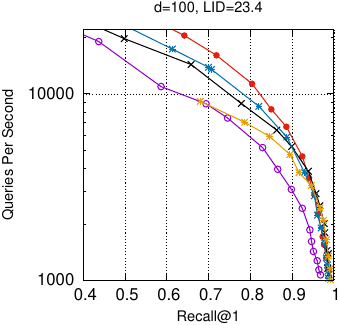}}
	\subfigure[GIST1M]
	{\includegraphics[width=0.186\linewidth]{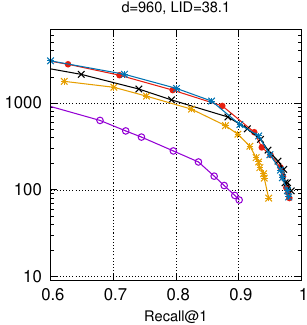}} 
	\hspace{0.01in} 
	\subfigure[GloVe1M]
	{\includegraphics[width=0.189\linewidth]{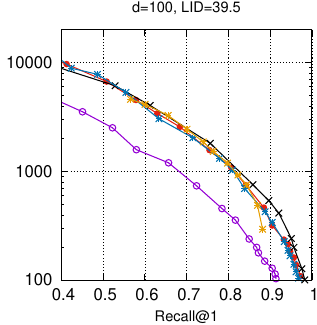}}
	\hspace{0.01in}
	\subfigure[Kosarak]
	{\includegraphics[width=0.185\linewidth]{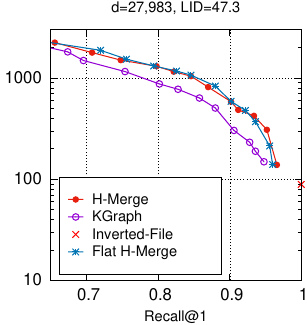}}
	\hspace{0.01in} 
	\subfigure[SIFT10M]
	{\includegraphics[width=0.191\linewidth]{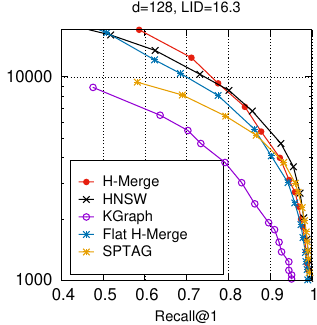}} 
 \caption{The NN search performance from representative graph-based approaches on ten datasets. The performance is given as the number of queries one approach could process in one second against Recall@1 on this speed level. The series of \#queries/recall pairs for each curve are produced by varying the critical parameter \textit{k} in the search. Parameter \textit{k} is the size of the top-rank list in the search procedure. Generally, larger \textit{k} leads to higher search quality while lower speed efficiency.}
\label{fig:nns}
\end{center}
\end{figure*}

\subsection{Performance on NN Search}
In this section, we study the performance of H-Merge as it is adopted to build the hierarchy graphs for the fast NN search. The search follows the pipeline described in Section~\ref{sec:srch}. In the hierarchy, five layers are kept for datasets except for Kosarak. The sizes of graphs in each layer (from top to bottom) are set to \textit{64}, \textit{512}, \textit{4,096}, \textit{32,768} and \textit{n}, where \textit{n} is the size of the whole reference set. For Kosarak, only four layers are kept. This setting is fixed all the time in the experiments. Ten datasets are used in the evaluation. They are one \textit{10} million, two \textit{1} million level synthetic datasets, and seven real-world datasets. The data dimension ranges from \textit{4} to \textit{27,983}. The reason to incorporate a series of synthetic datasets is to make the NN search performance trend more observable because their intrinsic data dimension increases steadily with respect to their data dimension. The general information is seen in Tab.~\ref{tab:datasets}. For all of the datasets except GloVe1M, NUSWide, and Kosarak, $l_2$ distance measure is used. For GloVe1M, we use \textit{Cosine} distance. $\kappa^2$ distance and \textit{Jaccard} distance are used for NUSWide and Kosarak respectively. In the study, we compare the search performance of H-Merge to recent graph-based approaches as well as the representative approaches of other categories.

\subsubsection{Comparison to Graph-based Approaches}
In the first evaluation, the performance of H-Merge is compared to other state-of-the-art graph-based approaches, namely HNSW~\cite{pami18:yury}, SPTAG~\cite{sptag}, and KGraph~\cite{weidong}. For KGraph, the approximate \textit{k}-NN graph is built by NN-Descent. An enhanced hill-climbing~\cite{icai11:kiana} procedure is adopted to perform the NN search over the approximate \textit{k}-NN graph. Each \textit{k}-NN graph in the graph hierarchy produced by H-Merge is diversified by Alg.~\ref{alg:gd}. In addition, in order to investigate the impact of hierarchical structure in H-Merge, another run for H-Merge is undertaken. In this run, the bottom layer graph that is produced by H-Merge and diversified by Alg.~\ref{alg:gd} is directly used for the NN search. The search procedure is basically the same as the second phase search (described in Section~\ref{sec:srch}) except that the seeds are randomly selected. This run is given as ``Flat H-Merge''. On datasets NUSWide and Kosarak, the performance from SPTAG is not reported because $\kappa^2$ distance and \textit{Jaccard} distance are infeasible for SPTAG. Instead, the performance from Inverted-File is reported on Kosarak as it is believed to be effective for the high dimensional sparse datasets.

For H-Merge and KGraph, the hyper-parameter \textit{k} is fixed to \textit{40} across all the datasets. On the one hand, this setting is recommended in~\cite{dpg:wenli}. On the other hand, our offline test confirms that this setting makes a good trade-off between the search quality and efficiency. In HNSW, parameters \textit{M} and \textit{efConstruction} are set to \textit{20} and \textit{128} respectively. Parameter \textit{M} is the allowed connections to each sample, which is comparable to the allowed neighbors in a diversified graph~\cite{dpg:wenli}. Since the allowed neighbors in our diversified H-Merge bottom graph are set to $k/2$ (\textit{i.e., 20}), the parameter settings in H-Merge and HNSW are comparable. An offline test on \textit{M} and \textit{efConstruction} with different settings also confirm that current settings are appropriate across different datasets. In SPTAG, most of the parameters except for \textit{TPTNumber} follow the default settings. Parameter \textit{TPTNumber} is set to \textit{8}, which improves search performance on most datasets while reducing index construction time.

The graph quality as well as the computation cost to build the approximate \textit{k}-NN graph for \textit{10} datasets are shown in Tab.~\ref{tab:graph_recall}. For all the approaches studied here, parameter \textit{k} is fixed to \textit{40}. For the all approaches, the graph construction is sped up by OpenMP on \textit{24} cores. In general, H-Merge takes roughly twice more time to produce the approximate \textit{k}-NN graph than NN-Descent. This is a little bit above the complexity bound we analyzed in Section~\ref{sec:conv} due to the computation overhead induced by additional operations. However, this extra cost brings us an extra bonus, namely the graph hierarchy. The graph quality of H-Merge and NN-Descent is similar in most of the cases. For HNSW and SPTAG, many true neighbors have been deliberately removed from the graph, their approximate \textit{k}-NN graph quality is therefore considerably lower than the other. Both HNSW graphs and SPTAG are primarily designed to support NN search instead of \textit{k}-NN graph construction. 

The search performance of four approaches is shown in Fig.~\ref{fig:nns}. The performance is given as the number of queries that one approach could process in one second against the Recall@1 on this speed level. It is clear to see H-Merge, HNSW, and Flat H-Merge outperform KGraph by a large margin on most of the datasets. Since the quality of graphs that support the NN search for KGraph and Flat H-Merge is similar, the performance gap between them largely owes to the graph diversification operation that is adopted in Flat H-Merge. H-Merge and HNSW show significantly superior performance over Flat H-Merge on the low dimensional datasets such as RAND10M8D due to the support of the hierarchy structure. This is the unique advantage of H-Merge and HNSW over other graph-based approaches such as~\cite{weidong,dpg:wenli,pr19:hcnng}. The performance of SPTAG is poorer than H-Merge and HNSW on most of the datasets.

However, the performance superiority achieved by these two approaches fades away as the data dimension rises up to \textit{32}. As the dimensionality grows, both hierarchical and non-hierarchical approaches get more likely trapped in the close neighborhood. The speedup that one approach could achieve is related to the intrinsic dimensionality of the dataset. The high speedup is achieved when the ratio between the data dimension and the intrinsic data dimension is high. This observation is in line with the findings in~\cite{GD19, dpg:wenli}. 

Among the two approaches that show the best performance, H-Merge outperforms HNSW considerably on datasets RAND10M8D, YFCC1M, NUSWide, and GIST1M while shows similar performance as HNSW on the rest. Compared to HNSW, the advantage of H-Merge is that it applies graph diversification on a complete approximate \textit{k}-NN graph. The structure in the NN list neighborhood is well-preserved. H-Merge is more attractive over HNSW for at least two reasons. First of all, H-Merge shows relatively better performance on various data types. Moreover, H-Merge is able to produce and maintain a high quality approximate \textit{k}-NN graph for an open set. It is particularly helpful for multimedia websites, where we should maintain a dynamic \textit{k}-NN graph to allow users to browse over similar photos and videos and support fast NN search in the meantime.

On the high dimensional sparse dataset (shown in Fig.~\ref{fig:nns}(i)), the performance from H-Merge and Flat H-Merge is still competitive with Inverted-File, which attains \textit{100\%} recall while with slightly lower speedup. We believe that Inverted-File will show more significant superiority over graph-based approaches when the data dimensionality reaches millions.

\subsubsection{Comparison to Approaches in the Literature}
\begin{figure}
\begin{center}
 \subfigure[Recall@1=0.8]
 {\includegraphics[width=0.455\linewidth]{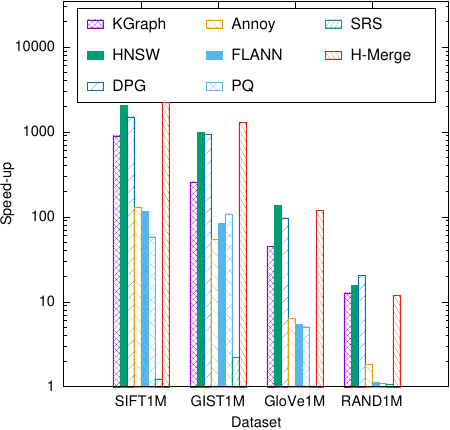}} 
 \hspace{0.15in}
 \subfigure[Recall@1=0.9]
 {\includegraphics[width=0.45\linewidth]{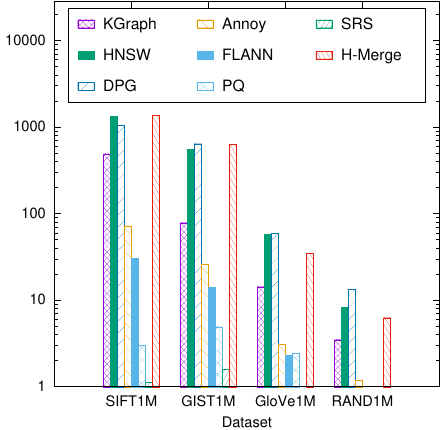}} 
 \caption{The performance of H-Merge in comparison to representative approaches in the literature on four million level datasets. In the evaluation, we study the speedup that one approach could achieve when the search quality is fixed to Recall@1=$0.8$ and Recall@1=$0.9$ respectively. The speedup is measured in contrast to the exhaustive search time cost shown in Tab.~\ref{tab:datasets}.}
\label{fig:all_nns}
\end{center}
\end{figure}
In this evaluation, we further study the performance of H-Merge in comparison to NN search approaches in the literature. Eight representative approaches are considered in the comparison. Namely, they are SRS~\cite{srs14}, PQ~\cite{JDS11}, FLANN~\cite{pami14:flann}, Annoy~\cite{annoy}, DPG~\cite{dpg:wenli}, KGraph~\cite{weidong} and HNSW~\cite{pami18:yury}. For DPG, its indexing graph is derived from the approximate \textit{k}-NN graph produced by NN-Descent. So it shares the same approximate \textit{k}-NN graph as KGraph. For all the approaches considered here, the parameters are set according to either the original paper or codes released by the authors. Four datasets that range from ``easy'' to ``hard'' are selected in the comparison. For each approach, we report its speedup over brute-force search when the top-\textit{1} search recall level is fixed at \textit{0.8} and \textit{0.9} respectively. Fig.~\ref{fig:all_nns} (a) and Fig.~\ref{fig:all_nns} (b) show the speedups on these two different levels.

As seen from the figure, graph-based approaches H-Merge, DPG, and HNSW outperform the approaches of other categories considerably. Compared to non-hierarchical graph-based approaches (DPG and KGraph), the performance from H-Merge and HNSW is overall slightly better. However, no significant performance gap is observed between hierarchical approaches (e.g., HNSW and H-Merge) and non-hierarchical approaches (e.g., DPG and KGraph) on these high dimensional datasets. This again confirms that the hierarchy structure is able to slightly alleviate the complexity induced by the increase of data dimension but unable to overcome it. The hierarchy structure is helpful when both the data dimension and the intrinsic data dimension are low. In general, HNSW and H-Merge perform similarly on these two snapshot of views.

In this paper, three graph merge algorithms are proposed. The theoretical convergence analysis of these algorithms (including NN-Descent) is also given. Although S-Merge, J-Merge, and H-Merge are built on top of NN-Descent, the problems that S-Merge, J-Merge, and H-Merge aim to address are new. NN-Descent alone is unable to solve the problems. They play different roles in the approximate \textit{k}-NN graph construction. S-Merge is suitable for parallel or distributed approximate \textit{k}-NN graph construction. J-Merge allows an approximate \textit{k}-NN graph to be incrementally built. From the practical point of view, we need all these approaches to be in place given the reminiscence of the big data problem from various fields. The original NN-Descent is only suitable for building a graph for a fixed dataset or building subgraphs for S-Merge and J-Merge use.

In addition to \textit{k}-NN graph construction, H-Merge is proposed to build a hierarchical indexing structure for fast NN search. It shows competitive performance with state-of-the-art approach HNSW. Compared to HNSW, a \textit{k}-NN graph is maintained. On the one hand, it facilitates the users to browse over NN search results via the links between neighbors in the task such as image/video infringement search. On the other hand, it is also the media that allows two indexing structures to be merged via S-Merge. In contrast, such functionalities are not available in HNSW.

\section{Conclusion}
\label{sec:conc}
We have addressed the issues of merging \textit{k}-NN graphs, which have been long overlooked in the literature. Three simple but effective solutions namely S-Merge, J-Merge, and H-Merge are proposed. They can be viewed as extensions over the classic NN-Descent algorithm and are tailored to addressing the issues that are unsolvable with the original NN-Descent. S-Merge is designed to merge two existing graphs, which is the critical step for parallel approximate \textit{k}-NN graph construction. J-Merge addresses the problem of building an approximate \textit{k}-NN graph for an open set, which is hardly achievable with state-of-the-art approaches. Deriving from J-Merge, H-Merge builds the approximate \textit{k}-NN graph in a hierarchical manner, which facilitates fast NN search, particularly for low dimensional data. Moreover, the hierarchical graphs built by H-Merge could grow incrementally as well with the support of J-Merge. All the merge algorithms presented in the paper maintain the coherence, genericness, and simplicity exhibited in NN-Descent. Extensive experiments have been carried out to verify the effectiveness of the proposed solutions both for approximate \textit{k}-NN graph construction and large-scale NN search.

\ifCLASSOPTIONcompsoc
 \section*{Acknowledgments}
\else
 \section*{Acknowledgment}
\fi

This work is supported by National Natural Science Foundation of China under grants 61572408 and 61972326, and the grants of Xiamen University 20720180074.

\bibliographystyle{ieeetr}
\bibliography{wlzhao}

\begin{IEEEbiography}[{\includegraphics[width=1in,clip,keepaspectratio]{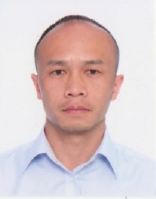}}]
{Wan-Lei Zhao} received his Ph.D degree from City University of Hong Kong in 2010. He received M.Eng. and B.Eng. degrees in Department of Computer Science and Engineering from Yunnan University in 2006 and 2002 respectively. He currently works with Xiamen University as an associate professor, China. Before joining Xiamen University, he was a Postdoctoral Scholar in INRIA, France. His research interests include multimedia information retrieval and video processing.
\end{IEEEbiography}

\vspace{-0.1in}

\begin{IEEEbiography}[{\includegraphics[width=1in,clip,keepaspectratio]{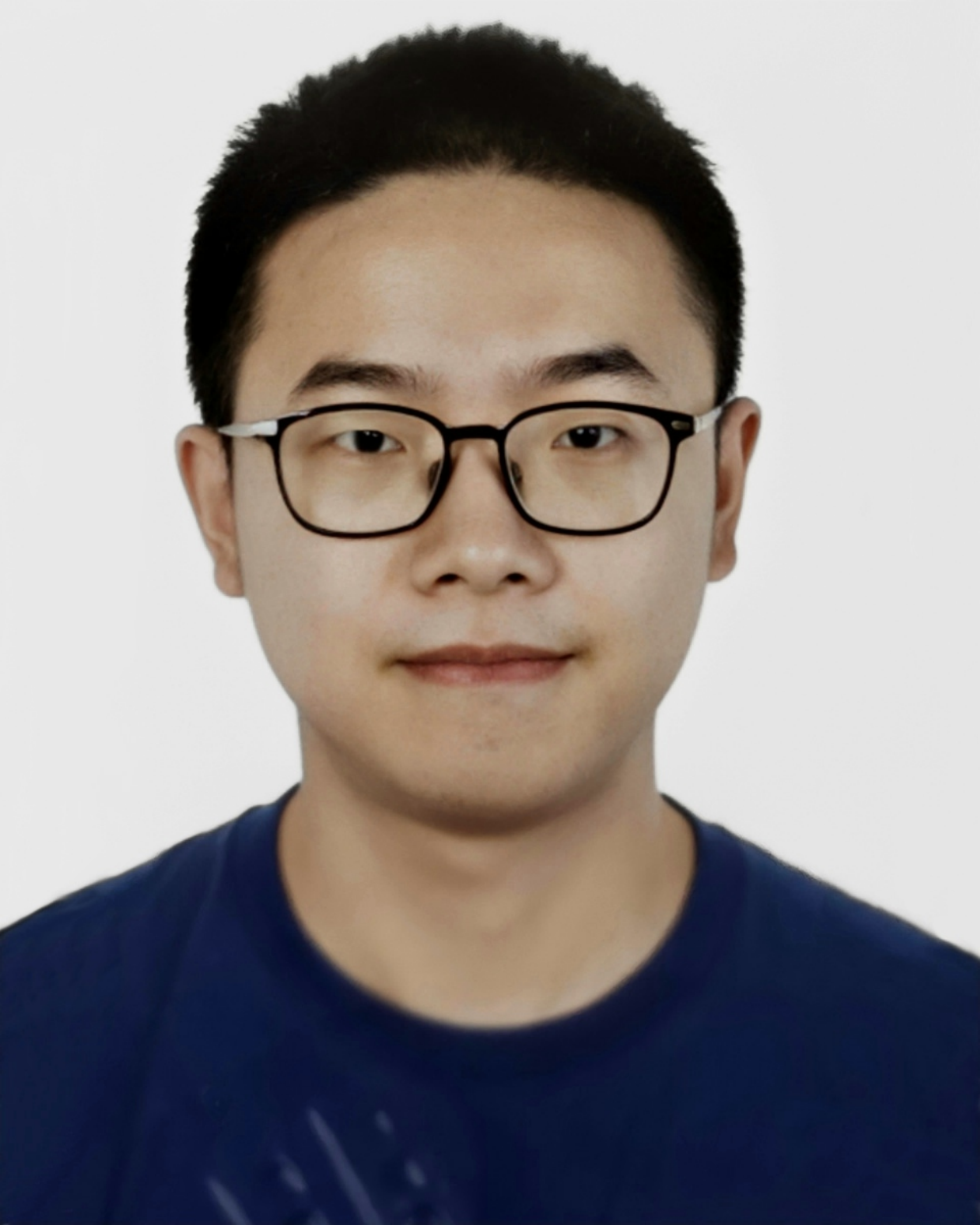}}]
	{Hui Wang} received his Bachelor degree of Engineering from Zhejiang Sci-Tech University, China in 2019. He is currently pursuing a Master’s degree in School of Informatics at Xiamen University. His research interest is large-scale nearest neighbor search.
\end{IEEEbiography}

\begin{IEEEbiography}[{\includegraphics[width=1.0in,clip,keepaspectratio]{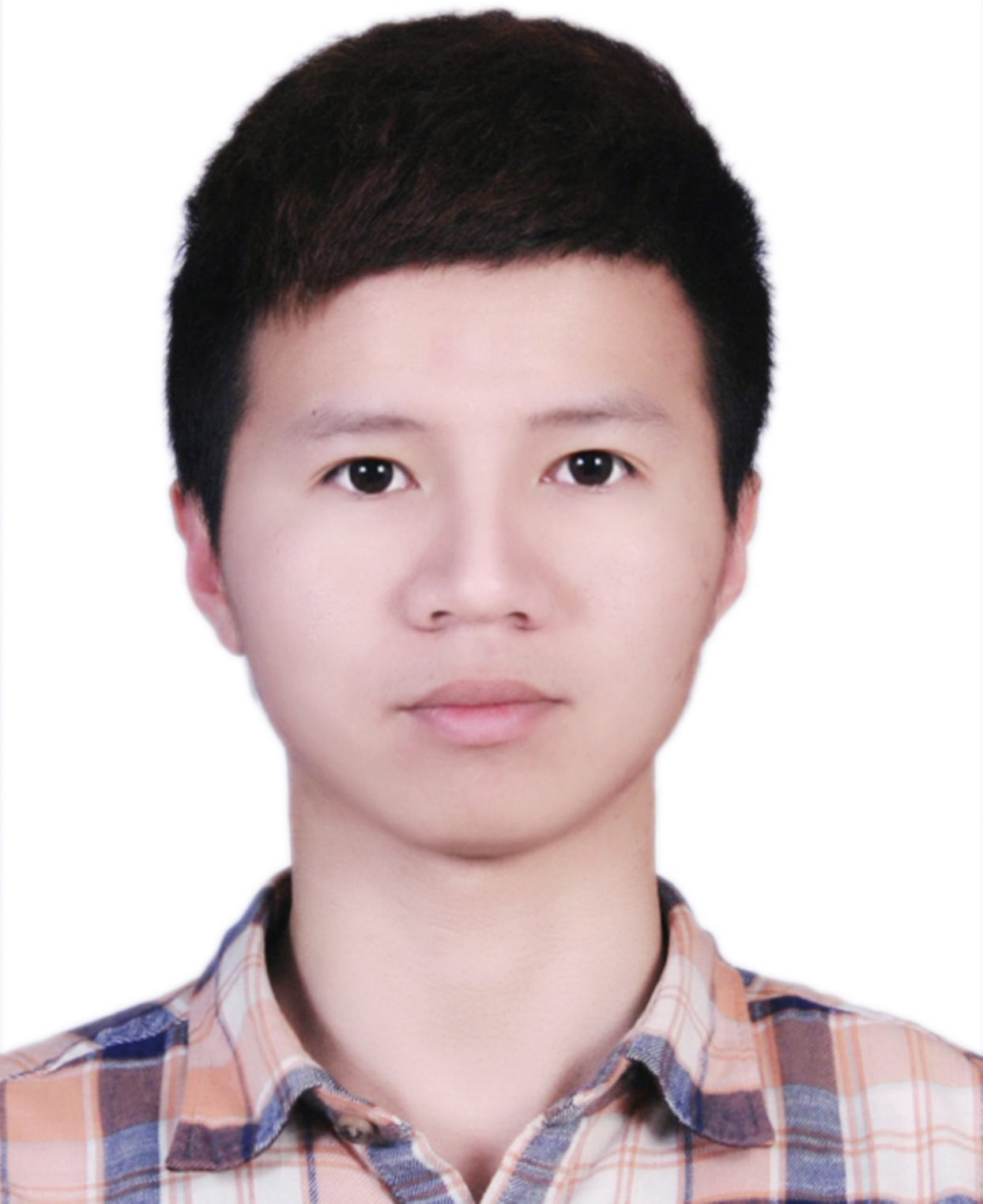}}]
{Peng-Cheng Lin} received his Bachelor degree of Engineering from Fujian Normal University, China in 2017. He is currently a graduate student at Department of Computer Science, Xiamen University. His research interest is large-scale nearest neighbor search. 
\end{IEEEbiography}


\begin{IEEEbiography}[{\includegraphics[width=1in,clip,keepaspectratio]{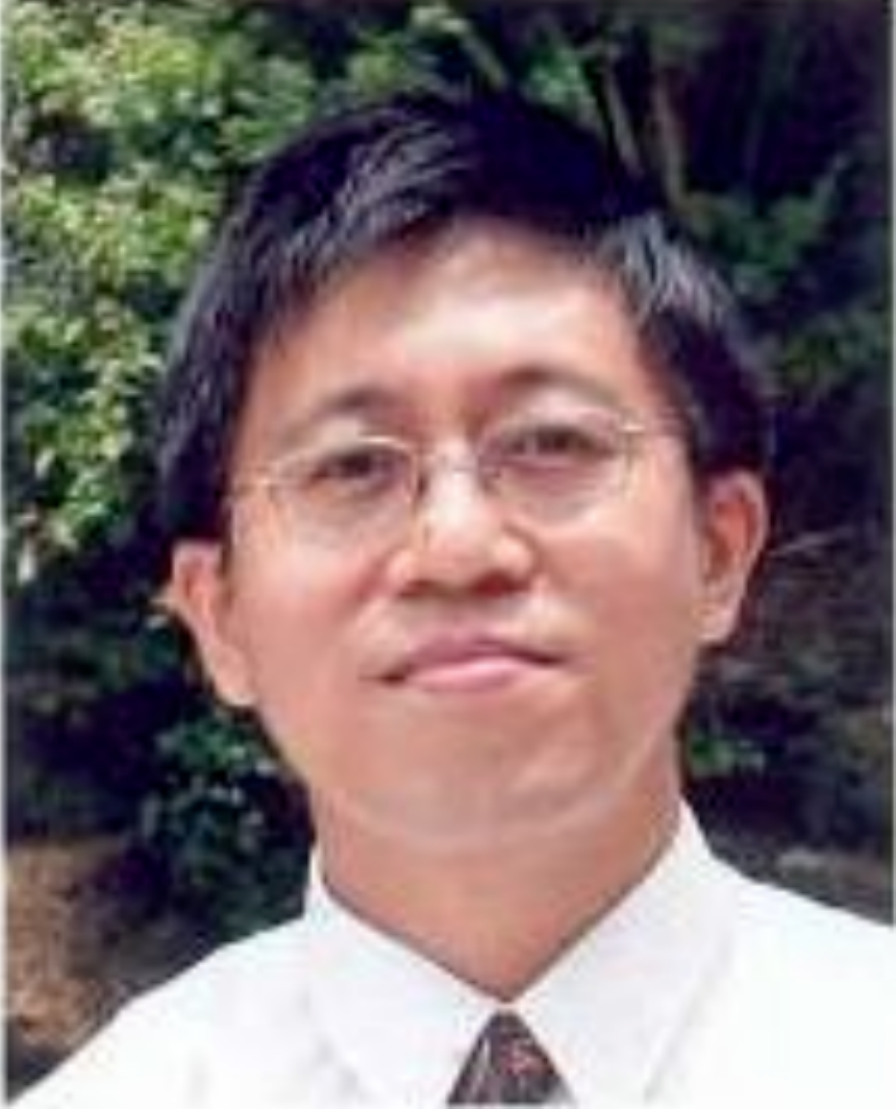}}]
{Chong-Wah Ngo} received the B.Sc. and M.Sc. degrees both in computer engineering from Nanyang Technological University of Singapore, Singapore, and the Ph.D. degree in computer science from Hong Kong University of Science and Technology, Hong Kong. He is a Professor with the Department of Computer Science, City University of Hong Kong, Hong Kong. Before joining City University of Hong Kong, he was a Postdoctoral Scholar with the Beckman Institute, University of Illinois, Urbana-Champaign. His main research interests include large-scale multimedia information retrieval, video computing, multimedia mining and visualization. He was an Associate Editor for the IEEE TRANSANCTION ON MULTIMEDIA and is currently steering committee member of TRECVid, ICMR (International Conference on Multimedia Retrieval) and ACM Multimedia Asia. He is program Co-Chair of ACM Multimedia 2019, and general Co-Chairs of ICIMCS 2018 and PCM 2018. He was named ACM Distinguished Scientist in 2016 for contributions to video search and semantic understanding.
\end{IEEEbiography}

\end{document}